\documentclass{book}
\usepackage{qpt_book}
\usepackage{amsfonts,amsmath,amssymb}
\usepackage{graphicx}
\usepackage{epstopdf}
\usepackage{wasysym}

\begin{document}

\pagestyle{myheadings}


\markboth{Understanding Quantum Phase Transitions}
{Metastable Quantum Phase Transitions in a One-Dimensional Bose Gas}

\chapter[Metastable Quantum Phase Transitions in a 1D Bose Gas]{Metastable Quantum Phase Transitions in a One-Dimensional Bose Gas}

\vskip -0.5cm

{\bf \large Lincoln D. Carr} \\
{\it Department of Physics, Colorado School of Mines, Golden, CO 80401, U.S.A.} \\
\vskip 2mm
\noindent
{\bf \large Rina Kanamoto} \\
{\it Division of Advanced Sciences, Ochadai Academic Production, Ochanomizu University,\\ Bunkyo-ku, Tokyo 112-8610 Japan} \\
\vskip 2mm
\noindent
{\bf \large Masahito Ueda} \\
{\it Department of Physics, University of Tokyo, \\Bunkyo-ku, Tokyo 113-0033 Japan} \\
\vskip 2mm

\noindent
This manuscript corresponds to chapter 13 of the book ``Understanding Quantum Phase Transitions", edited by Lincoln D. Carr (Taylor $\&$ Francis, Boca Raton, FL, 2010)

\vskip 4mm
\noindent
Ultracold quantum gases offer a wonderful playground for quantum many-body physics~\cite{leggett2001}, as experimental systems are widely controllable, both statically and dynamically.  One such system is the one-dimensional (1D) Bose gas on a ring.  In this system binary contact interactions between the constituent bosonic atoms, usually alkali metals, can be controlled in both sign and magnitude; a recent experiment has tuned interactions over seven orders of magnitude~\cite{09:Hulet}, using an atom-molecule resonance called a Feshbach resonance.  Thus one can directly realize the  Lieb-Liniger Hamiltonian (LLH)~\cite{63:L,63:LL}, from the weakly- to the strongly-interacting regime.  At the same time there are a number of experiments utilizing ring traps~\cite{03:Dem}.  The ring geometry affords us the opportunity to study topological properties of this system as well; one of the main properties of a superfluid is the quantized circulation in which the average angular momentum per particle, $L/N$, is quantized under rotation~\cite{leggett1999}. Thus we focus on a tunable 1D Bose system for which the main control parameters are interaction and rotation.  We will show that there is a critical boundary in the interaction-rotation control-parameter plane over which the topological properties of the system change.  This is the basis of our concept of \textit{metastable quantum phase transitions} (QPTs).  Moreover, we will show that the finite domain of the ring is necessary for the QPT to occur at all because the zero-point kinetic pressure can induce QPTs, i.e., the system must be finite; we thus seek to generalize the concept of QPTs to inherently finite, mesoscopic or nanoscopic systems.\footnote[1]{See also Chap. 27 for a discussion of finite-size effects on QPTs in the context of nuclei.}

Specifically, we will show that past the critical boundary the phase of a Bose-Einstein condensate (BEC) can wind and unwind continuously and can therefore no longer be considered a superfluid in the usual sense, despite a single mode still being macroscopically occupied~\cite{penrose1956}.  Since the interaction and rotation will be in units related to the ring size, another way to look at this critical boundary is that the superfluidity in this example is a mesoscopic effect and might vanish in the thermodynamic limit.  Using mean-field theory in the form of the celebrated Gross-Pitaevskii or nonlinear Schr\"odinger equation (NLSE), the presence of the critical boundary can be tied to the appearance of new stationary solutions, \emph{dark solitons},\index{soliton} which bifurcate from a \textit{uniform superflow} or plane-wave branch.  Dark solitons, which take the form of a density notch in BECs, can have zero density, in which case they are called \textit{black} and have a node, or a region of decreased density without a node, in which case they are called \textit{gray}.  Gray solitons have a characteristic phase structure which allows for the topological winding and unwinding, as we will show. As the density contrast of a gray soliton tends to zero, i.e., as it becomes shallower, its phase structure, density, and energy approach those of uniform superflow.  Arrays of regularly spaced dark solitons are called \textit{dark-soliton trains}.  Dark solitons, black and gray, singly and in trains, have been observed in BEC experiments, beginning with~\cite{burger1999,denschlag2000} and, more recently, in~\cite{weller2008,stellmerS2008}.   Although dark-soliton trains are robustly stable in quasi-1D mean-field theory,\footnote[2]{Quasi-1D means that the mean field is confined but the underlying binary scattering problem remains 3D.  See~\cite{dunjko2001} for details.} in practice they are found to decay in experiments.  This has been attributed first to use of 3D geometries, in which dark soliton nodal planes decay via the snake instability into vortex--anti-vortex pairs; in this case mean-field theory provides an excellent description of the ensuing dynamics. However, even in quasi-1D experiments dark solitons are found to decay; the main cause is thought to be thermal~\cite{jackson2006} and/or quantum fluctuations~\cite{03:DKS,carr2009}.

The study of quantum fluctuations, and the use of dark-soliton decay as a smoking-gun signal for fluctuations, has a long track record.  It turns out that dark solitons can also be observed in optical fibers with anomalous dispersion and a Kerr nonlinearity~\cite{kivshar1998}.  The governing equation is again the NLSE, and dark solitons decay due to periodically spaced fiber amplifiers.  This effect is called Gordon-Haus jitter~\cite{gordon2}, and can be formulated as a 1D random walk with the accompanying diffusion equation.  In BECs, thermal effects can play a similar role~\cite{fedichev1999,jackson2006}.  However, at very low temperatures and for a quasi-1D mean-field one is left with only quantum fluctuations as the cause of dark soliton decay.  Thus a thorough investigation of such fluctuations was undertaken by Dziamarga \textit{et al.}, starting with~\cite{03:DKS}.  Using the Bogoliubov-de Gennes equations (BdGE) and modified version thereof, they found that although dark solitons delocalized, due to the anomalous or Goldstone mode, they did not in fact decay.  That is to say that averaging over many measurements one finds the soliton fills in, but for a single measurement a dark soliton is found somewhere in the condensate.  However, the BdGE are limited to weak interactions.  Recently one of us studied this problem in the more strongly interacting regime, placing the soliton in an optical lattice to enhance quantum fluctuations and using matrix-product-state (MPS) algorithms to follow the dynamics.\footnote[3]{Specifically, we used time-evolving block decimation(TEBD); see Chap.~23.}  We found that a dark soliton does indeed decay at zero temperature, based on three main pieces of evidence: density-density correlations, inelasticity in soliton-soliton collisions, and a close comparison to Bogoliubov theory via an analysis of modes of the single-particle density matrix.  It has recently been demonstrated that it is possible in the weakly interacting regime for the filling-in of density-density correlations to be consistent with delocalization rather than decay~\cite{martinAD2009,deuar2010}; however, the strongly-interacting regime remains an open question.  Thermal fluctuations in dark solitons have also been used to study the dynamics of phase transitions in the Kibble-Zurek mechanism~\cite{damski2009}; see Chap.~3.

We will show that dark-soliton trains on a ring are in fact the weakly-interacting limit of \textit{yrast states}~\cite{99:BR, 99:BP}, and that it is really the yrast states that play the key role in our finite-size metastable QPT when analyzed in the full many-body problem beyond mean-field theory.
Yrast, a Swedish term originally used in nuclear physics which can be translated as
`dizziest,' refers to the lowest energy state for a given angular momentum.  We will identify Lieb's \textit{Type II} or hole excitations with the yrast states, solving the long-standing question of the physical interpretation of this solution branch of the LLH~\cite{80:IT}.  The uniform superflow states which the soliton trains bifurcate from will be replaced with the more general {\it center-of-mass rotation} (CMR) states, a special class of yrast states for which $L/N$ is an integer. In the NLS mean-field theory, the phase winding number comes out of a calculation of the circulation; out of the mean-field regime, such terminology becomes questionable if not meaningless.  Thus the whole concept of uniform superflow and of superfluidity breaks down; yet our QPT persists as an accurate picture of an abrupt transition in even the strongly-interacting system's properties.  In fact, we can take this description all the way to the \textit{Tonks-Girardeau} (TG) limit in which one obtains an exact solution to the problem via the TG Bose-Fermi mapping~\cite{yukalov2005}.  At the end of our chapter, we will come back to the connection between yrast states and quantum fluctuations of dark solitons.

Finally, we wish to make a brief comment on dimensionality.  Although there is formally no superfluid phase transition in 1D, nevertheless for an interacting finite system there is in practice still a sharp crossover to a macroscopically occupied mode at a critical temperature.  One should keep in mind that bounded and/or finite domain reduced-dimensional systems often have distinctly different properties from their infinite counterparts, for instance graphene, which is unstable on the infinite domain due to thermal fluctuations, but, as we know now, not only stable on finite domain but has many important technical applications~\cite{geim2007}.  Likewise, this is an exciting time for the study of 1D systems, not so much for the possibility of revolutionizing computing as with graphene, but for revolutionizing our understanding of quantum many-body physics.  1D systems have some special advantages for that purpose.  First, there are established exact solution methods such as the Bethe ansatz for special but nevertheless important systems, including the focus of the present study, the LLH.  We will take advantage of the plethora of 1D theoretical techniques in our study in order to bring to bear multiple lines of evidence on our problem.  Second, experiments in ultracold quantum gases~\cite{04:KWW} are allowing for the thorough investigation and testing of fundamental 1D questions, including Kolmogorov-Moser-Arnold (KAM) theory for integrable quantum systems, quantum quenches, and a host of other zero and finite-temperature effects.  Third, there are exciting new developments in MPS numerical methods that allow one to follow entangled quantum many-body dynamics of 1D systems, integrable or nonintegrable, as discussed in Chaps.~22 and~23.

This chapter draws strongly from three papers~\cite{08:KCU,09:KCU,10:KCU}, the foundations for which were laid by~\cite{05:KSU}; however, our previous work on dark solitons~\cite{00:CCR,carr2000e,carr2001e,carr2009} also plays an important role in our discussion and our thinking.  Our set of theoretical techniques for this study included the NLSE, the BdGE, exact diagonalization in a truncated angular momentum basis, the finite-size Bethe ansatz equations, and the TG equations.\footnote[4]{An MPS approach is also possible, but requires spatial discretization, an aspect we wish to avoid.  A continuum limit of MPS is possible in principle but tricky, requiring typically thousands of lattice sites; see~\cite{muth2009}.}  We will sketch the results of these techniques, referring the reader to~\cite{09:KCU,10:KCU} for details.

Unlike the excited state QPTs in nuclear physics described in Chap.~27, our QPT occurs in an inherently finite system. The QPT here disappears in the thermodynamic limit and cannot be extrapolated from finite size scaling or other such arguments.  Thus there is no concept of nonanalyticity (see, e.g., Chap.~1) and it is formally a crossover.  Given the prevalence of nano-devices and other inherently finite systems, we think the term \textit{finite-system} QPT is a useful one, as we hope to convince the reader in the course of this chapter.

\section{Fundamental Considerations}\index{Lieb-Liniger Hamiltonian}
\label{sec:LLH}

We begin with a clear statement of the Hamiltonian.  In position representation the LLH is~\cite{63:LL}
\begin{equation}\label{LLH}
\hat{H}_0 = - \textstyle\sum_{j=1}^N \partial_{\theta_j}^2
+ g_{\rm 1D}\textstyle\sum_{j<k}\delta(\theta_j-\theta_k),
\end{equation}
where $\theta_j$ is the azimuthal angle that satisfies $0\le \theta_j < 2\pi$ and $g_{\rm 1D}$ is the binary contact interaction strength renormalized for our 1D problem.  We take the experimental context of the LLH as $N$ bosons in a thin torus, or ring trap.  The torus has radius $R$; we take the length, angular momentum, and energy
to be measured in units of $R$, $\hbar$, and $\hbar^2/(2 m
R^2)$, respectively, throughout our treatment.

In a rotating frame of reference with an angular frequency $2\Omega$ the LLH becomes
\begin{equation}\label{rotH}
\hat{H}(\Omega)=\hat{H}_0 - 2\Omega \hat{L} + \Omega^2N\,,
\end{equation}
where
\begin{equation}
\hat{L}\equiv -i\textstyle\sum_{j=1} \partial_{\theta_j}
\end{equation}
is the angular-momentum operator; we refer to Eq.~(\ref{rotH}) as the \textit{rotating LLH} (rLLH).
From the single-valuedness boundary condition of the many-body wave function~\cite{73:Legg},
one can show that solving the eigenproblem in the rest frame $\hat{H}_0 \Psi_0 = E(0) \Psi_0$ suffices
in order to obtain solutions to the eigenproblem $\hat{H}(\Omega)\Psi=E(\Omega)\Psi$~\cite{09:KCU}.
The eigenvalue is then obtained from
$E(\Omega)=E(0)-2\Omega \langle \hat{L} \rangle +\Omega^2 N$, which is periodic with respect
to $\Omega$.

The subclass of solutions to the rLLH we focus on are the yrast states.
This approach is useful for our ring system, because all the information about physical properties in
the rotating frame are embedded within the spectrum in the rest frame.
Thus the physical meanings of yrast states can be extracted by the simple
transformation of yrast spectra.
Since the LLH commutes with the angular-momentum operator, $[\hat{H}_0,\hat{L}]=0$,
the yrast problem is well defined irrespective of the sign and strength of interaction,\footnote[5]{Throughout this chapter we focus on repulsive interactions $g_{\rm 1D}>0$, for which QPTs only occur in excited states.  However, for attractive interactions, $g_{\rm 1D}<0$, QPTs occur also in the ground state~\cite{05:KSU}.}
and all the yrast states are eigenstates of both
the LLH $\hat{H}_0$ and the rLLH $\hat{H}(\Omega)$.

We can identify the yrast states by dividing the solution space of $\hat{H}_0$ and $\hat{H}$ into subspaces according to the two conserved quantities in the problem: the number of bosons $N$, and the total angular momentum $L\equiv\langle\hat{L}\rangle$.  In each
such subspace we can index excited states by $q\in\{1,2,\ldots\}$ in ascending order in energy; thus the state-ket is written $|N,L; q\rangle$.
Then the yrast states are denoted as
$|N,L; q=1\rangle$.
The essential properties of the ground and low-lying excited states can be described
within the yrast states, as we have verified explicitly with exact diagonalization studies.\footnote[6]{Such studies necessarily used a truncated single-particle angular-momentum basis, as the Hilbert space is otherwise too large.}  Thus we henceforth omit the quantum number $q$
from the notations for eigensolutions.
With the abbreviation of the quantum number $q=1$ and
for fixed coupling constant $g_{\rm 1D}$,
the eigenvalues that correspond to the yrast states are written as
$E_{N,L}(\Omega)$, where we explicitly write the parameter $\Omega$ in the notation
in order to clarify in which frame the system is.  With this notation, the eigenvalues
in the nonrotating frame are written as $E_{N,L}(0)$.

Most of the yrast states will turn out to be dark solitons for weak interactions, but in this limit a special subclass corresponds to uniform superflow.  Due to the translational invariance of the LLH with respect to $\theta$ and $\Omega$,
properties of a particular set of yrast states can be analyzed without solving
the problem.  These are the CMR states.  In particular, they are the states for which the total angular momentum
is equal to an integer multiple of the total number of atoms $N$.
The energy of the CMR state takes the form
\begin{equation}\label{sf_ene}
E_{N,L=JN}(0)=J^2 N +V_{\rm int}\,,
\end{equation}
where $V_{\rm int}$ is the interaction energy and $J \in \mathbb{Z}$ is an integer.
We call $J$ the {\it center-of-mass quantum number}, because
it physically expresses the amount of uniform translation of the center-of-mass momentum.
In the NLS mean-field theory, $J$ is conventionally called
the phase winding number, but this is a valid concept only in the weakly interacting limit, and we retain the more general notion for our purpose of identifying the QPT through all interaction regimes.
In the rotating frame, the energy of the CMR state is given by
\begin{equation}\label{sf_ene_rot}
E_{N,JN}(\Omega)=(J-\Omega)^2N+V_{\rm int}\,,
\end{equation}
where the change in energy associated with the frame change is involved only
in the kinetic energy term, and the interaction energy
is completely separated from the parameter $\Omega$.

The ground state in the absence of the
rotating drive is the state with zero angular momentum, $E_{N,L=0}(0)$.
The excitation energy of the CMR states with a finite angular momentum $L=JN$ is thus given by
\begin{equation}\label{sf_ex_ene}
E_{N,JN}(0)-E_{N,0}(0)=J^2 N,
\end{equation}
which is independent of the strength of interaction $g_{\rm 1D}$. This is natural
because changing the total angular momentum by the amount $JN$ leads to a Galilean transformation without changing the boundary condition for the order parameter.
The ground state in the presence of the rotating drive is characterized
by the CMR quantum number\footnote[7]{$\lfloor x  \rfloor$ is the floor function, meaning the largest integer that does not exceed $x$.}
\begin{equation}\label{CMqn}
J_0=\lfloor \Omega+1/2 \rfloor.
\end{equation}

Because of the periodicity in the eigensolutions, an eigenstate $|N,L\rangle$ with
the energy $E_{N,L}(\Omega)$ has
a denumerably infinite number of counterparts {$|N,L+JN\rangle$} and $E_{N,L+JN}(\Omega)$,
corresponding to arbitrary values of $J \in \mathbb{Z}$.
Solving the yrast problem for a limited range of fixed angular-momentum states,
e.g., $-N/2 \le L < N/2$, therefore suffices to obtain all the eigensolutions.
Moreover, the spectra are degenerate for the same magnitude of angular momentum,
$E_{N,L}=E_{N,-L}$ in the absence of rotating drive, while this degeneracy is
resolved in the presence of rotation due to the Sagnac effect~\cite{13:Sag}.
All other yrast states for $L$ out of this limited range can be obtained by
shifting the total angular momentum
by $N$ while keeping the internal structure of the eigenstates.
This is similar to a band theory concept~\cite{09:KCU}, with $-N/2 \le L < N/2$ playing the role of the first Brillouin zone.

\section{Topological Winding and Unwinding: Mean-Field Theory}\index{topological winding}\index{mean field theory}
\label{sec:NLSE}

\begin{figure}[t]
\begin{center}
\includegraphics[width=0.8\textwidth]{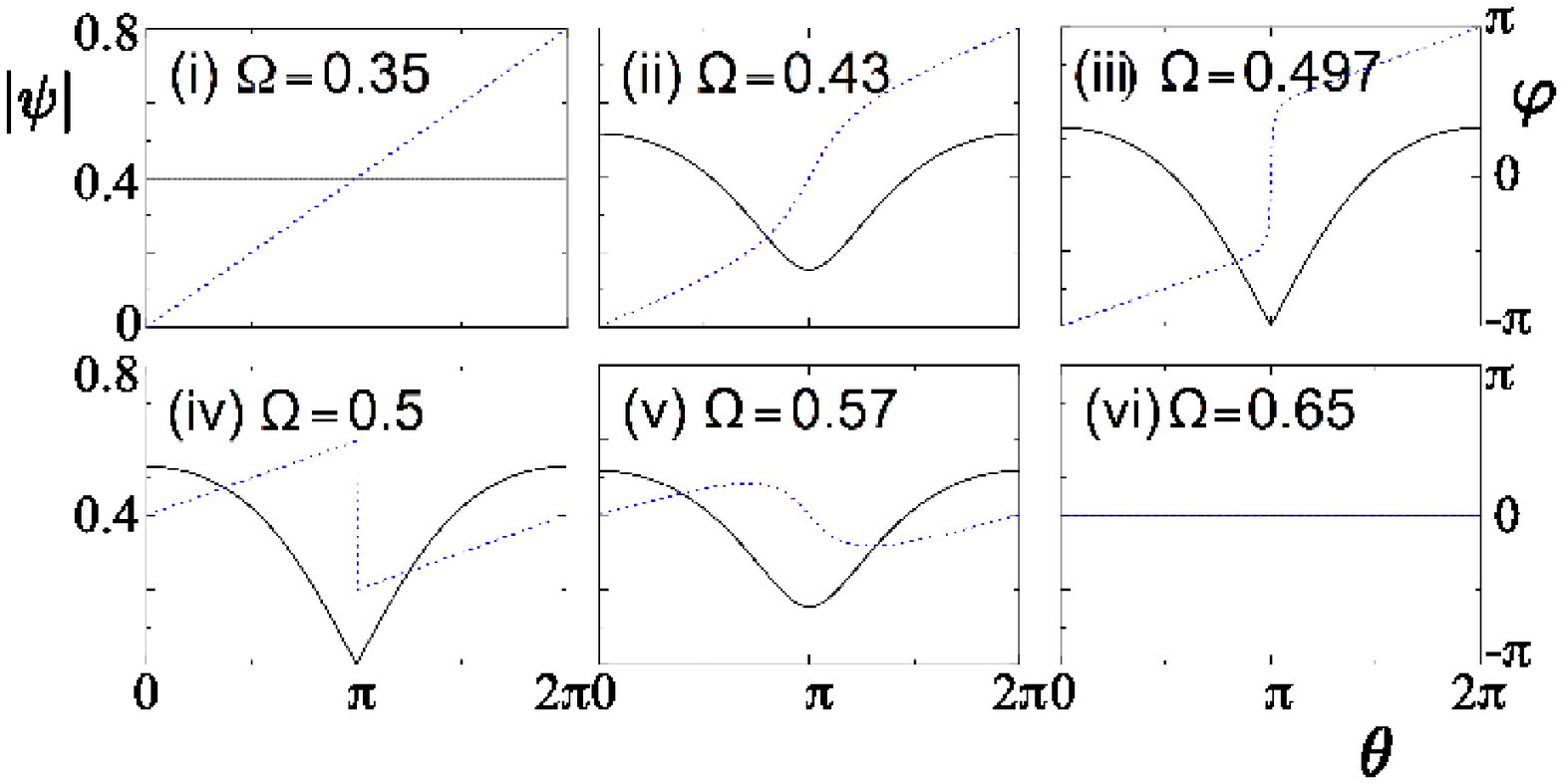}
\vspace*{-0.5cm}\end{center}
\caption{
Topological winding and unwinding of a BEC via formation of a dark soliton.  Shown are the mean-field amplitude (solid curves with the left reference) and
phase (dotted curves with the right reference).  Uniform solutions with different values
of the phase winding (i) $J=1$ and (vi) $J=0$ are smoothly connected
through the broken-symmetry dark soliton (ii)--(v)
with a self-induced phase slip at (iv) $\Omega=0.5$.  Reproduced from~\cite{08:KCU}.
\label{CKU:fig1}}
\end{figure}

We first introduce the basic concept of the finite-size metastable QPT in the simplest context, namely mean-field theory.  We restrict ourselves to a fairly qualitative discussion; a complete quantitative theory, including full analytical solutions and perturbative stability analysis, can be found in~\cite{08:KCU,09:KCU}.

The NLSE, which can be derived directly from the rLLH~\cite{leggett2001},\footnote[8]{The key assumption is to take a macroscopically occupied mode in the single particle density matrix and neglect small fluctuations around that mode; subsequently one replaces averages over products of operators with the product over averages.} is
\begin{equation}
[(-i\partial_{\theta}-\Omega)^2+g_{\rm 1D}N|\psi(\theta)|^2]\psi(\theta)=\mu\psi(\theta),
\end{equation}
where all bosons are taken to be in the same macroscopic mode which is described by the order parameter $\psi$; this order parameter can be interpreted physically in terms of a density, $|\psi|^2$, and a phase, $\varphi\equiv \mathrm{Arg}(\psi)$. The single-valuedness of the wave function requires
$\varphi(\theta+2\pi)=\varphi(\theta)+2\pi J$, where $J$ is an integer and
can, in this weakly-interacting limit, be understood as a topological winding number.

The uniform superflow solutions of the NLSE are just plane waves, $\psi(\theta)=\psi_0\exp(iJ\theta)$, while the dark-soliton train solutions are expressed in terms of Jacobi elliptic functions.\footnote[9]{In the infinite-system limit a single dark soliton is written in terms of a tanh function.}
We take the number of density notches in the dark-soliton train to be $j$.
For the rest of this section, we consider the single soliton $j=1$ for simplicity,
but our discussion holds for arbitrary soliton trains $j>1$.

Figure~\ref{CKU:fig1} illustrates how an initial uniform-superflow solution with $J=1$ can be continuously unwound to $J=0$.  As $\Omega$ increases starting
from (i) the uniform superflow with $J=1$,
(ii) solitons start to form past a critical point $\Omega_{\mathrm{crit}}^{(1)}$.
(iii) The density notch deepens for $\Omega_{\rm crit}^{(1)} \le \Omega \le 0.5$.
(iv) At $\Omega=0.5$ it forms a node, the phase of the soliton jumps by $\pi$,
and the energies of the solitons with phase winding number 1 and 0
are degenerate. (v) The soliton with phase winding $J=0$ deforms continuously
as $\Omega$ increases. (vi) Finally, the state goes back to
the uniform-superflow state with phase winding $J=0$. For the purposes of illustration we have taken a fixed interaction strength $g_{\mathrm 1D} N =0.6$; in fact, interaction and/or rotation can be used to wind or unwind the order parameter; likewise, one can characterize the whole process in terms of system size, while holding other parameters fixed.

In Fig~\ref{CKU:fig1} we observe the characteristic phase structure of dark solitons: inside the density notch, the phase changes more rapidly, and the total difference across the notch yields the soliton's velocity; outside the density-notch region, the phase has a uniform slope, indicating the velocity of the uniform superflow.  Since a dark soliton moves with a velocity which is opposite to the direction of increase in phase (to the left in Fig.~\ref{CKU:fig1}), one can understand such stationary solutions as a cancelation between dark-soliton motion to the left, clockwise looking down on the ring from above, and uniform superflow to the right, or counter-clockwise.

\begin{figure}[t]
\begin{center}
\includegraphics[width=0.7\textwidth]{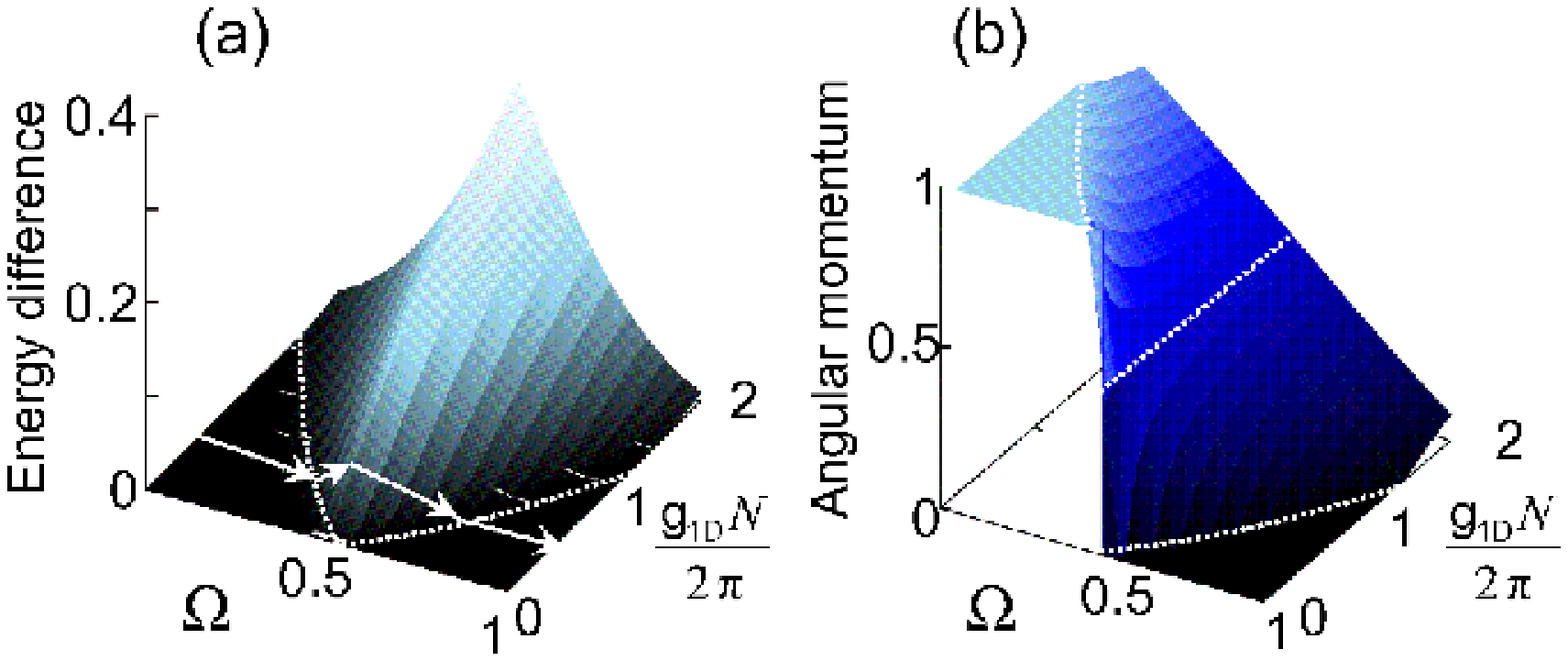}
\vspace*{-0.5cm}\end{center}
\caption{
(a) The bifurcation between uniform-superflow and dark-soliton states are illustrated in terms of (a) energy difference and (b) average angular momentum per particle $L/N$.  The soliton solutions only exist in the area surrounded by two critical boundaries (white dotted curves).  The white arrow in (a) indicates the path taken during the unwinding of the phase shown in Fig.~\ref{CKU:fig1}.  In (b) one clearly sees two topologically distinct regions: in between the critical boundaries the characteristic quantization of $\bar{L}$ in the BEC breaks down, even though there is still macroscopic occupation of a single mode.  Reproduced from~\cite{08:KCU}.
}
\label{CKU:fig2}
\end{figure}

Figure~\ref{CKU:fig2}(a) shows the energy difference
between a dark soliton and uniform superflow, as can be calculated analytically.
This kind of bifurcation does not occur from the ground-state energy.
However, for metastable states a bifurcation can occur between the uniform-superflow state and the soliton state with the same winding number $J$.
After bifurcation, the soliton energy becomes
larger than the uniform-superflow energy.  For convenience, Fig.~\ref{CKU:fig2}(a) displays a white arrow indicating the higher-energy, soliton path taken in Fig.~\ref{CKU:fig1}.  Exact diagonalization studies were used to show that an arbitrary potential which breaks the symmetry of the ring automatically sets the system on this path.  Shown in Fig.~\ref{CKU:fig2}(b) is the key physical observable for our QPT, the average angular momentum per particle, expressed as $L/N=\int d\theta \psi^*(-i\partial_{\theta})\psi$, in the mean-field theory.
Thus a continuous
change of angular momentum is possible for 1D Bose systems
by taking the metastable states with energy slightly higher than
that of the ground state.  Both panels of Fig.~\ref{CKU:fig2} display a lower and an upper critical angular frequency, which we call $\Omega_{\mathrm{crit}}^{(1)}$ and $\Omega_{\mathrm{crit}}^{(2)}$, as a function of $g_{\rm 1D}$.

Bifurcation of the soliton train from the uniform superflow constitutes
a second-order QPT with respect to $g_{\rm 1D}$ and/or $\Omega$.  The derivatives of the soliton and uniform-superflow energies with respect to $\Omega$ have a kink at
the boundary as can be verified analytically. This identifies
the QPT, at least at the mean-field level, which
occurs along a curve in the $\Omega$-$g_{\rm 1D}$ plane.  We also found that the Hessian for the energy is discontinuous  along this curve, while the Hessian for the chemical potential diverges.

\section{Finding the Critical Boundary: Bogoliubov Analysis}\index{Bogoliubov-de Gennes equations}\index{critical boundary}
\label{sec:BdGE}

We can better understand the QPT through the BdGE.  The BdGE can be derived formally as the lowest order quantum fluctuations from the LLH; it can also be simply thought of as linear perturbation theory on the NLSE, and is taken as such in optics.  In either case, without Bogoliubov analysis in terms of the BdGE we cannot be certain that our dark-soliton train solutions are stable even in the mean-field limit.  Moreover, the BdGE allow us to explicitly identify the critical boundaries shown in Fig.~\ref{CKU:fig2}.

A stationary solution $\psi(\theta)$ of the NLSE under a small perturbation $\delta$
evolves in time as
\begin{equation}
\tilde{\psi}(\theta,t)=e^{-i\mu t}\{\psi(\theta)+\sum_n [\delta u_n(\theta) e^{-i\lambda_nt}+\delta v_n^*(\theta) e^{i\lambda_n^* t}]\},
\label{eqn:BdGE}
\end{equation}
where $(u_n, v_n)$, and $\lambda_n$ are the eigenstates and eigenvalues of
the BdGE, respectively,
and $n\in|\mathbb{Z}|$ denotes the energy-index of excitations.
Recalling the structure of solutions in the Bogoliubov formalism~\cite{01:FS},
for each eigenvalue $\lambda_n$ with positive norm,
\begin{equation}
\textstyle\int_0^{2\pi}d\theta\,\left[|u_n(\theta)|^2-|v_n(\theta)|^2\right]=1,
\label{BdGnorm}
\end{equation}
there is also an eigenvalue $\bar{\lambda}_n\equiv -\lambda_{n}$ with negative norm.
The BdGE predict an infinite set of such solutions. One exception to this rule can exist.
This exception corresponds to \emph{Nambu-Goldstone modes}, sometimes called anomalous modes.
For instance, consider a black soliton is at rest on the ring with respect to any background superflow
in the rotating frame.  The Goldstone mode of the soliton corresponds to
the soliton moving at a small constant velocity in this frame.
We need consider only the Goldstone mode and eigenstates that satisfy~(\ref{BdGnorm}), since the eigenvalues
with negative norm do not have physical meaning.
For the Goldstone mode the corresponding eigenvalue is zero, and the latter eigenstates
that have positive norms can be real (i.e., positive or negative) or complex depending
on the stability of the condensate mode.  In general, $\lambda_n\in\mathbb{C}$.

The excitations of a uniform superflow with phase-winding $J$ are straightforward to calculate~\cite{01:FS}.  Fluctuations from that condensate mode are
given by the eigensolutions of the BdGE with positive norm,
\begin{equation}\label{pw_bog}
\lambda_l^{(J,{\rm us})}=\sqrt{l^2 (l^2+2\gamma)}-2l(\Omega-J), \quad u_l\propto e^{i(J+l)\theta},\quad v_l\propto e^{-i(J-l)\theta},
\end{equation}
where the `us' subscript stands for uniform superfluid, $\gamma\equiv g_{\mathrm{1D}}N/2\pi$, and $l\in\mathbb{Z}$
denotes the single-particle angular momenta of the excitation, a good quantum number since $[\hat{H}(\Omega),\hat{L}]=0$.

For $\gamma >0$, all the eigenvalues $\lambda_l^{(J,{\rm us})}$ are real,
as apparent from Eq.~(\ref{pw_bog}).  From Eq.~(\ref{pw_bog}) we also find that
several negative eigenvalues, associated with eigenstates of positive norm,
appear when we take a metastable {\it excited} state as a condensate mode.
These negative eigenvalues correspond to other plane-wave branches located in lower energy regimes
than the input condensate mode itself.
For the case of repulsive interactions, the number of negative eigenvalues thus coincides with the
number of stationary states that are located in a lower energy regime than the metastable state under consideration.  Solutions of the BdGE for dark-soliton trains require a numerical calculation.

We can identify the critical boundary $(\gamma_{\rm crit},\Omega_{\rm crit})$ for the QPT by finding where a particular negative eigenvalue changes its sign.
One equates $l$ to be $\pm j$ in Eq.~(\ref{pw_bog}) and
imposes the condition on the eigenvalue, $\lambda_{l={\cal S}j}^{(J,{\rm us})}=0$.  Then the critical boundaries are given implicitly by
\begin{eqnarray}\label{boundary}
\Omega_{\rm crit}-J=\pm\sqrt{(j/2)^2+\gamma_{\rm crit}/2}\,,
\end{eqnarray}
where the $\pm$ sign gives the lower and upper boundary, respectively, which we called $\Omega_{\rm crit}^{(1)}$ and $\Omega_{\rm crit}^{(2)}$.
The region $\gamma \le \gamma_{\rm crit}$ is identical to requiring that
the critical-boundary condition has a real solution.

\begin{figure}[t]
\begin{center}
\includegraphics[scale=0.45]{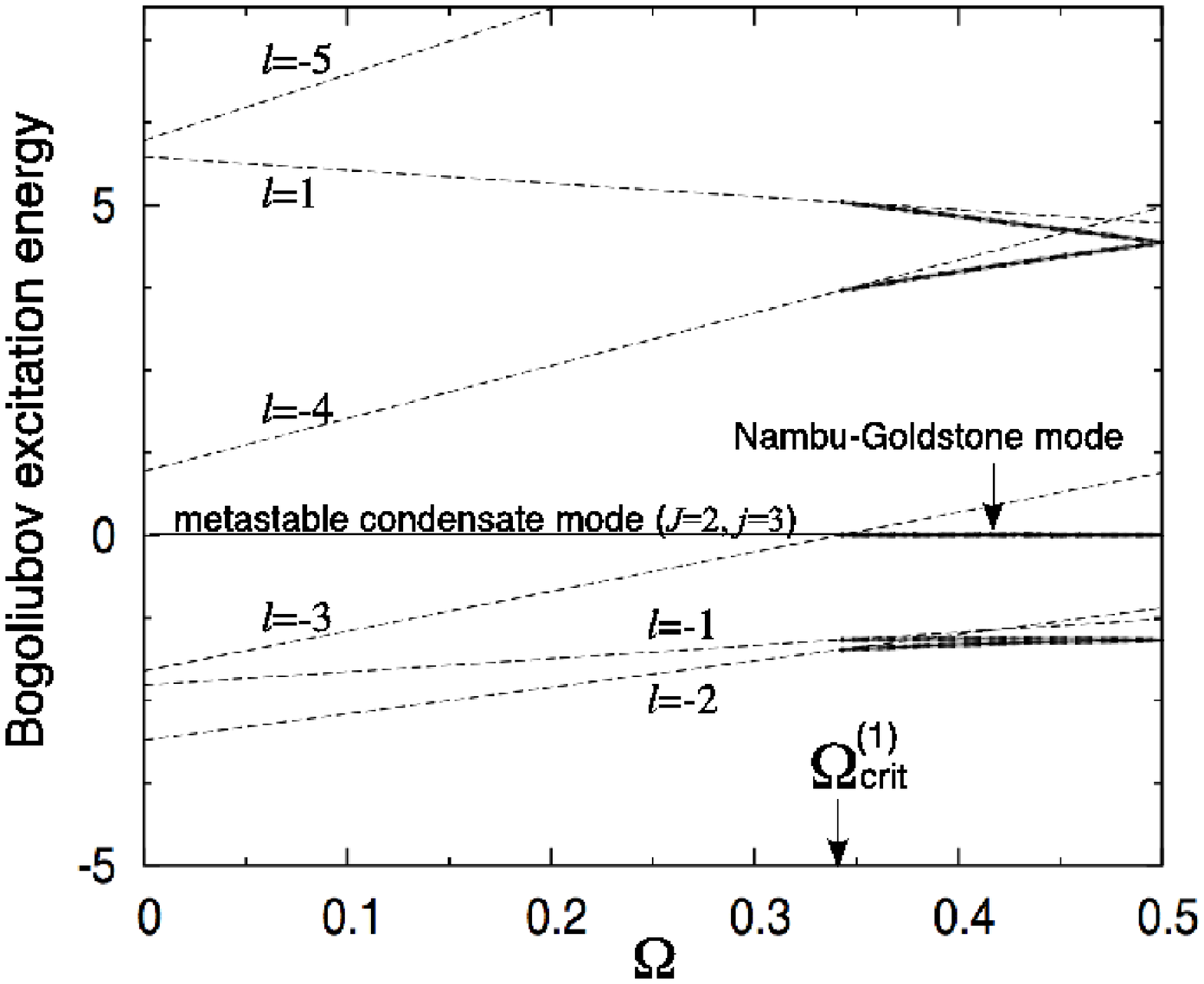}
\end{center}
\caption{Bogoliubov analysis of uniform superflow (thin dashed curves) and dark-soliton trains (thick solid curves).
The initial uniform-superflow state (zero excitation energy shown as the thin solid curve and labeled with text) with phase winding $J=2$ becomes the Nambu-Goldstone mode of the three-notch ($j=3$) dark-soliton train, past the critical frequency $\Omega_{\mathrm{crit}}^{(1)}$.
Other curves show excitation energies from the uniform and soliton train states, respectively, with various angular momenta labels $l$.  We take fixed interaction $g_{\rm 1D}N/2\pi = 1$ for this illustration. Reproduced from~\cite{09:KCU}.
}
\label{CKU:fig3}
\end{figure}
In Fig.~\ref{CKU:fig3} we illustrate the QPT from uniform-superflow to dark-soliton solutions; we illustrate only $0\leq \Omega\leq 0.5$ since the eigenvalues are symmetric for the other half of the Brillouin zone.  The soliton-train solutions occurring for $\Omega > \Omega_{\rm crit}^{(1)}$ are therefore linearly stable.\footnote[1]{They are also nonlinearly stable~\cite{carr2000e}.}
The excitation energies from the soliton branch are found to be close to those from the plane-wave branch.
The notable feature in the soliton regime is that there appears a Nambu-Goldstone mode,
which is continuously connected with one of the negative eigenstates with $l={\cal S}j$ from the plane wave.
This mode reflects the \textit{spontaneous symmetry breaking} of the soliton-train state.
At the point $\Omega_{\rm nodes}=0.5$, a degenerate pair of excitation branches emerges, where the phase jumps up or down by $\pi$ at each soliton in the soliton train.  This degeneracy is similar to that seen for a single soliton in Fig.~\ref{CKU:fig1}, but slightly complicated by the multiple solitons and higher phase winding.
For $\Omega > \Omega_{\rm nodes}$ the excitation branches from the soliton train are also symmetric with respect to
$\Omega_{\rm nodes}$.

For expediency we skip more details of BdGE solutions and summarize the overall results.  We found that excitations
from the uniform superflow in the $j^{\rm th}$ excited-state have $j$ thermodynamically unstable modes within the critical region in the $(\gamma,\Omega)$-plane.  In $\Omega_{\rm
crit}^{(1)} < \Omega \le 0.5$ in the dark-soliton train states become stable.
At the critical boundary, all energies of the stationary solutions and
the eigenvalues of the BdG equations continuously connect to the excitations in
the soliton regime without any energy discontinuity.
In particular, when one of the excitation energies from a plane-wave metastable state
changes sign from negative to positive, a soliton branch with the same phase-winding number
appears, and the Nambu-Goldstone mode manifests as a result.

Thus, upon consideration of general phase winding $J$ and soliton-number $j$, one realizes that there are a denumerably infinite set of paths to connect uniform-superflow states
via soliton trains in a metastable system of scalar bosons on a ring.  Associated with this transition, the energy of the solitons bifurcates, and a continuous change in the angular momentum becomes possible in the mean-field theory past a critical boundary in the interaction-rotation plane.  In general, bifurcations can indicate symmetry breaking associated with a QPT, as in the elementary example of the quadratic-quartic potential, where a single potential well bifurcates into two as the relative strengths of quadratic and quartic terms are tuned.



\section{Weakly-Interacting Many-Body Theory: Exact Diagonalization}\index{exact diagonalization}
\label{sec:ED}

We now show how in general to distill
the uniform-superflow and soliton-train mean-field branches from a sea of many-body eigenvalues.
We show how the mean-field soliton branch, for which average angular momentum
is not quantized as an integer, emerges from the yrast spectra. The meaning of spectra related to
symmetry breaking and the Nambu-Goldstone mode associated with the existence of the soliton branch is also discussed.

In order to take the next step towards a microscopic many-body solution of the LLH, it is convenient to rewrite it in second-quantized form.  One expands the bosonic field operator in terms of a plane-wave basis with the single-particle
angular momentum $l$,
\begin{equation}
\hat{\psi}(\theta)=(2\pi)^{-1/2}\textstyle\sum_{l=-\infty}^{+\infty} \hat{b}_l \,e^{il\theta}\,,
\label{eqn:discretization}
\end{equation}
where the pre-factor of $(2\pi)^{-1/2}$ comes from the normalization of the plane wave,
and $\hat{b}_l$ and $\hat{b}^{\dagger}_l$ are annihilation and creation operators
which obey the usual
commutation relations for bosons. Equation~(\ref{eqn:discretization})
satisfies the periodic boundary condition $\hat{\psi}(\theta)=\hat{\psi}(\theta+2\pi)$.
Then one finds
\begin{equation}\label{Ham2}
\hat{H}_0 = \textstyle\sum_{l=-\infty}^{+\infty}l^2\hat{b}_l^{\dagger}\hat{b}_l
+g_{\rm 1D}\textstyle\sum_{k,l,m,n=-\infty}^{+\infty}
\hat{b}_k^{\dagger}\hat{b}_l^{\dagger}\hat{b}_m\hat{b}_n\delta_{k+l,m+n}\,.
\end{equation}
\begin{figure}[t]
\begin{center}
\includegraphics[width=0.4\textwidth]{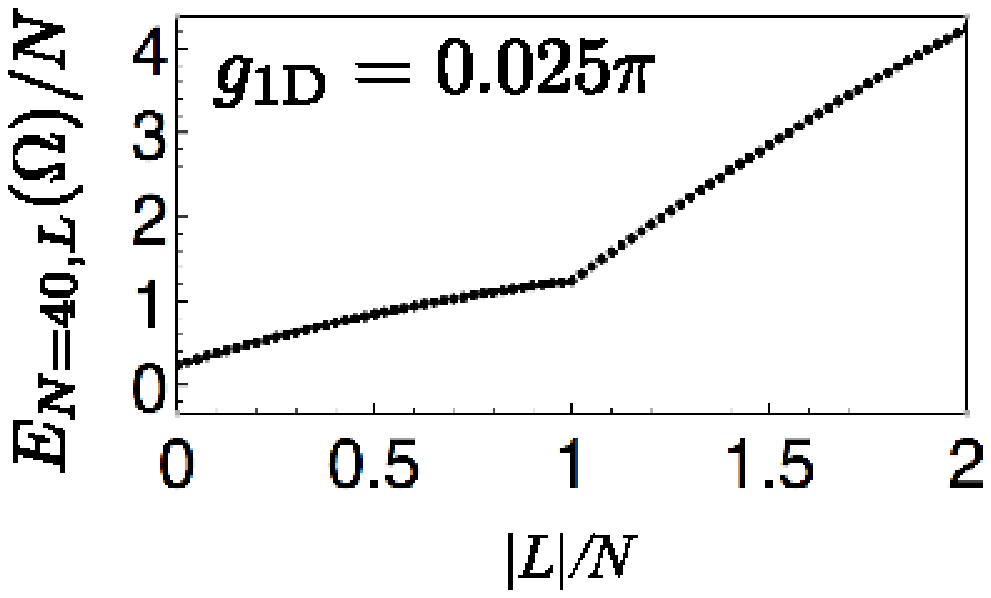}
\vspace*{-0.5cm}\end{center}
\caption{Yrast energy eigenstates for $N=40$ bosons on the ring, obtained by exact diagonalization of the rLLH with the cutoff
angular momentum $|l_c|=2$.
The spectrum has a kink where $L$ is an integer multiple of $N$.  Everything is symmetric for $L<0$, and in the large $N$ limit, the number of points increases and the discrete yrast energies
approach a continuous curve while the curvature and kink points remain unchanged. Reproduced from~\cite{10:KCU}.
}
\label{fig1}
\end{figure}
\noindent
The eigenstates can be expanded in terms of a Fock-state basis $|\{n_l\}\rangle$ that
represents the occupation number of each single-particle angular-momentum state,
\begin{equation}
|\{n_l\}\rangle=|\ldots, n_{-1}, n_0, n_1,\ldots\rangle\,.
\end{equation}
These states satisfy the conservation laws
\begin{equation}
\textstyle\sum_l n_l = N,\quad \textstyle\sum_l ln_l = L\,.
\end{equation}
For numerical calculations we use a cutoff angular momentum
$l_{\rm c}\geq 0$; thus $L \in [-l_{\rm c} N,l_{\rm c}N]$.
In the weakly interacting regime $g_{\rm 1D}N \lesssim \mathcal{O}(1)$,
a cutoff of $l_{\rm c}=2$ provides a quantitative
agreement in energy eigenvalues\footnote[2]{The convergence is better for (i) a larger number of atoms,
(ii) lower eigenvalues, and (iii) smaller strength of interaction.
Figure~\ref{fig7} indicates that for $N=10$ and up to the yrast states $|L|=2N$,
the deviation of the exact diagonalization results from the
Bethe ansatz results starts to emerge for $g_{\rm 1D} \gtrsim 1 $.
The results for $N=40$ in this section has, in general, better
agreement with the rigorous results.}
with those obtained by the Bethe ansatz shown in Sec.~\ref{sec:BA}.

Figure~\ref{fig1} shows the yrast energies for the nonrotating LLH,
$E_{N,L}(\Omega=0)=\langle N,L|\hat{H}_0|N,L \rangle$, for interaction strength $g_{\rm 1D}=2.5\times 10^{-2}\pi$ and number of bosons $N=40$.
The ratio of the mean-field interaction energy to the kinetic energy
corresponding to these values of $g_{\rm 1D}$ and $N$ is $g_{\rm 1D}N/(2\pi)=0.5$.
The kink observed in the figure is exactly the CMR state $L=JN$, here $L=N$.  In the mean field, this would be the uniform-superflow solution.  The other non-CMR yrast states contribute to the dark-soliton solutions, as we will show.

\begin{figure}[t]
\begin{center}
\includegraphics[width=0.8\textwidth]{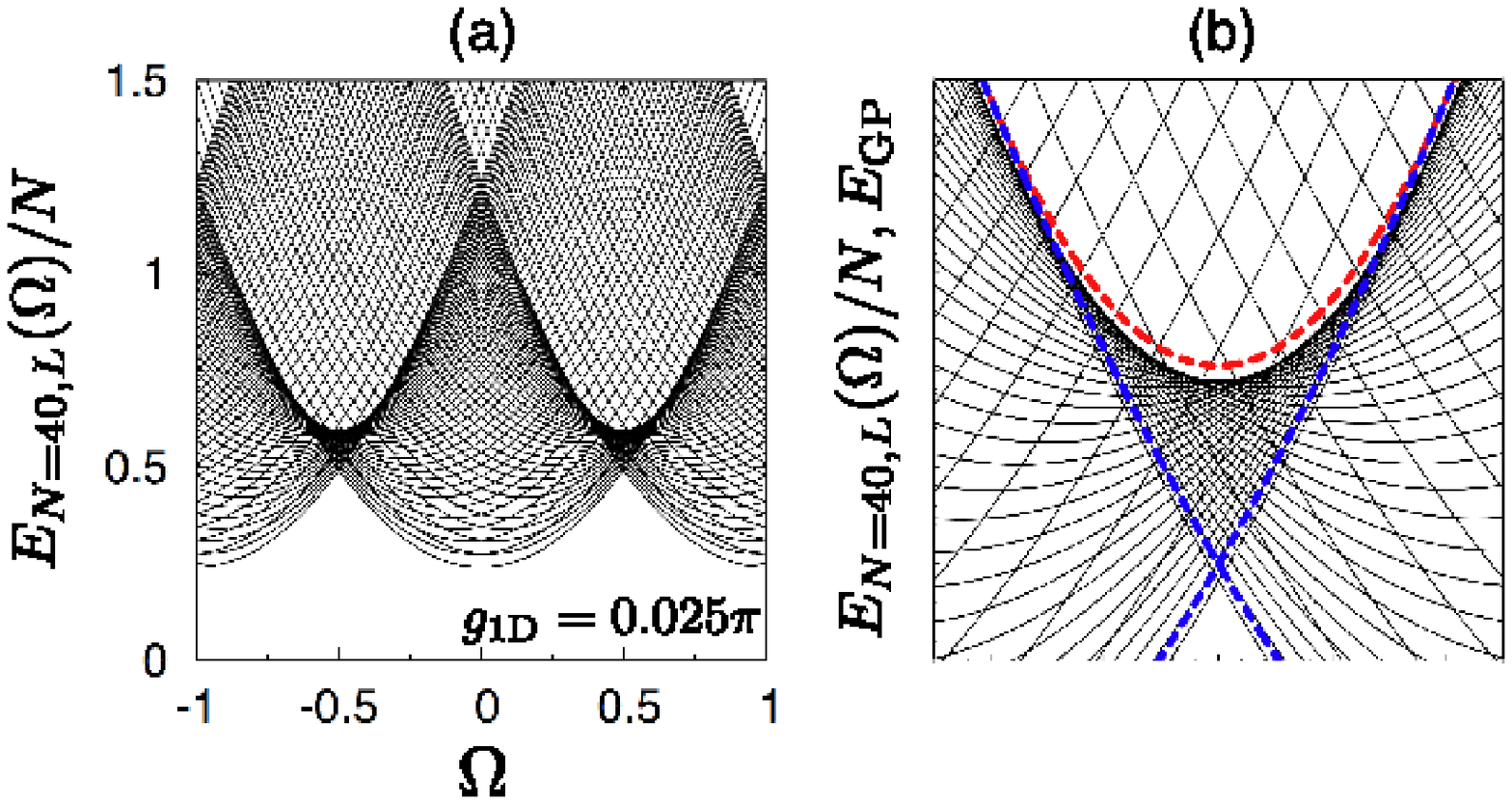}
\vspace*{-0.5cm}\end{center}
\caption{
(a) Performing a Legendre transform on the yrast states in Fig.~\ref{fig1}, we obtain the full yrast spectrum of the rLLH as a function of $\Omega$.
Each curve is distinguished by a different total angular momentum $L$.
(b) A zoom of (a) in the important region where solitons appear in the mean-field theory.
The dashed curves show a comparison to the NLSE superflow (blue) and soliton (red) branches.  The enclosed region is called a swallowtail.  Reproduced from~\cite{10:KCU}.
}
\label{fig2}
\end{figure}

To generalize Fig.~\ref{fig1} to the rotating rLLH, all we have to do is follow the procedure laid out in Sec.~\ref{sec:LLH}.  We perform a Legendre transform on the LLH yrast spectrum,
\begin{equation}
E_{N,L}(0)\to E_{N,L}(\Omega)=E_{N,L}(0)-2\Omega L + \Omega^2N.
\label{Lege}
\end{equation}
Figure~\ref{fig2} plots transformed yrast energies $E_{N,L}(\Omega)$, where
the finite number of points from Fig.~\ref{fig1}, each of which was characterized by a different
angular momentum $L$, become convex downward curves in Fig.~\ref{fig2}.
Each curve is thus characterized by a different total angular momentum and has a minimum
at a certain value of $\Omega$.

The energy $E_{N,L=J_0N}(\Omega)$, shown in Fig.~\ref{fig2}, corresponds to the ground state where
$J_0$ is the ground-state CMR quantum number given by Eq.~(\ref{CMqn}).
The angular-momentum states with $L=JN$ correspond to
the CMR states, and the center of the parabola is located at
$\Omega\in \{ \mathbb{Z} \}$ at which the CMR state becomes the ground state.

In Fig.~\ref{fig2}(a) we observe an extremely high density of states around $\Omega\in \{\pm 0.5, \pm 1.5,\dots\}$
due to the crossing of many eigenvalues, as shown more clearly in the zoom in panel (b).  The region of high density takes the shape of a \textit{swallowtail}; the same shape was found to occur purely within mean-field theory, past the critical boundary for the QPT~\cite{09:KCU}.
This swallowtail region is almost filled by various energy eigenvalues of
various angular-momentum states crossing each other.
The domain with the high-density swallowtail shape looks as if it is
enclosed by the two kinds of stationary branches predicted by the mean-field theory.

How can we understand this spectrum?  It is an interesting fact that mean-field theories tend to break the symmetries of many-body theories.  In our case, if the NLSE has a soliton solution, the solution must be localized on the ring; the NLSE is nonlinear and cannot work with superpositions of solitons.  In contrast, the quantum theory finds a superposition of all possible positions.\footnote[3]{There is a quantum entropy associated with this degeneracy~\cite{00:CCR}.} This is the main reason why Dziamarga \textit{et al.} found that an initial mean-field soliton delocalizes at a higher level of quantum theory~\cite{03:DKS}.

We can also see this effect in vortex formation in a scalar condensate under rotation.
Solving the yrast problem in 2D results in all the angular-momentum states,
including the at-rest condensate ($L=0$), off-axis vortex ($0 < L < N$),
a centered vortex ($L=N$), and
vortex lattices ($L > N$). However, in experiments one drives the system with a
specific angular frequency. In such a situation, there exists a small distortion
in the trap, which selects a metastable angular-momentum state with respect to
the variation in the angular momentum of the condensate. As a result, in reality
one does not observe a stationary off-centered vortex except as a transient state.

\begin{figure}[t]
\begin{center}
\includegraphics[width=0.4\textwidth]{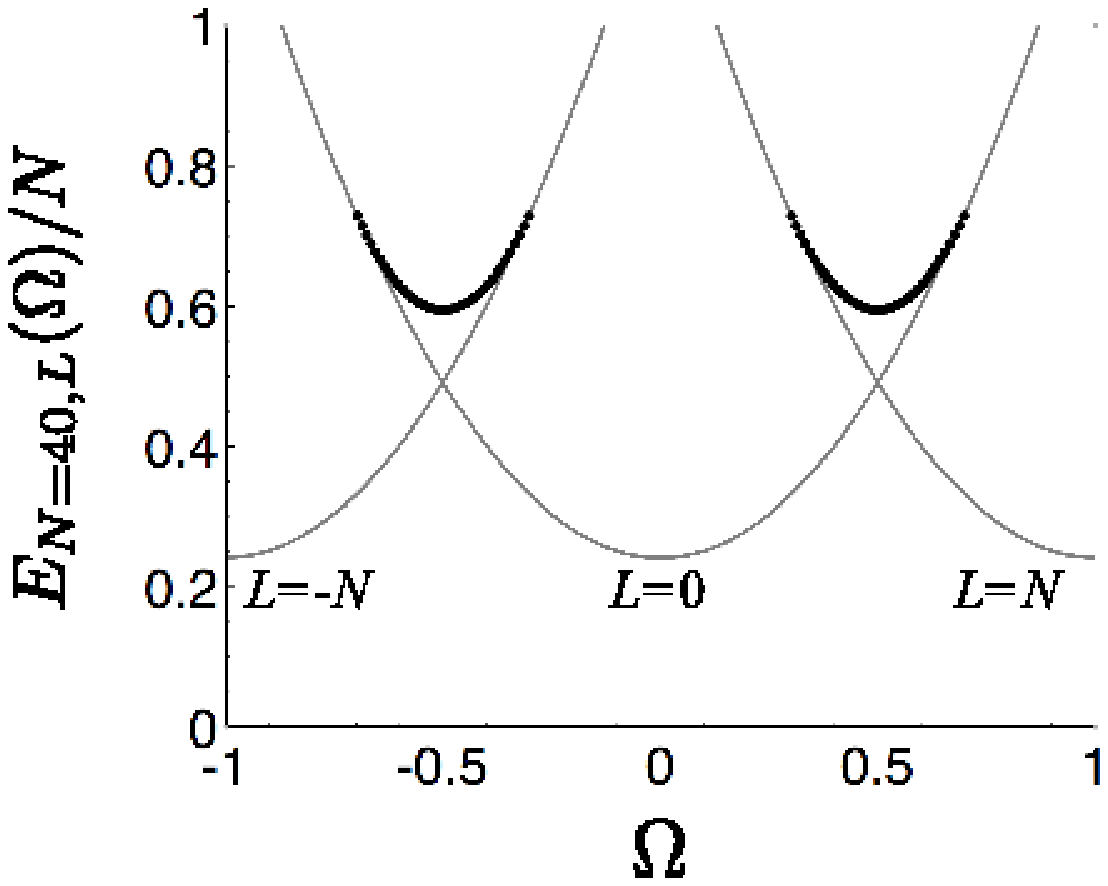}\hspace*{0.05\textwidth}
\includegraphics[width=0.4\textwidth]{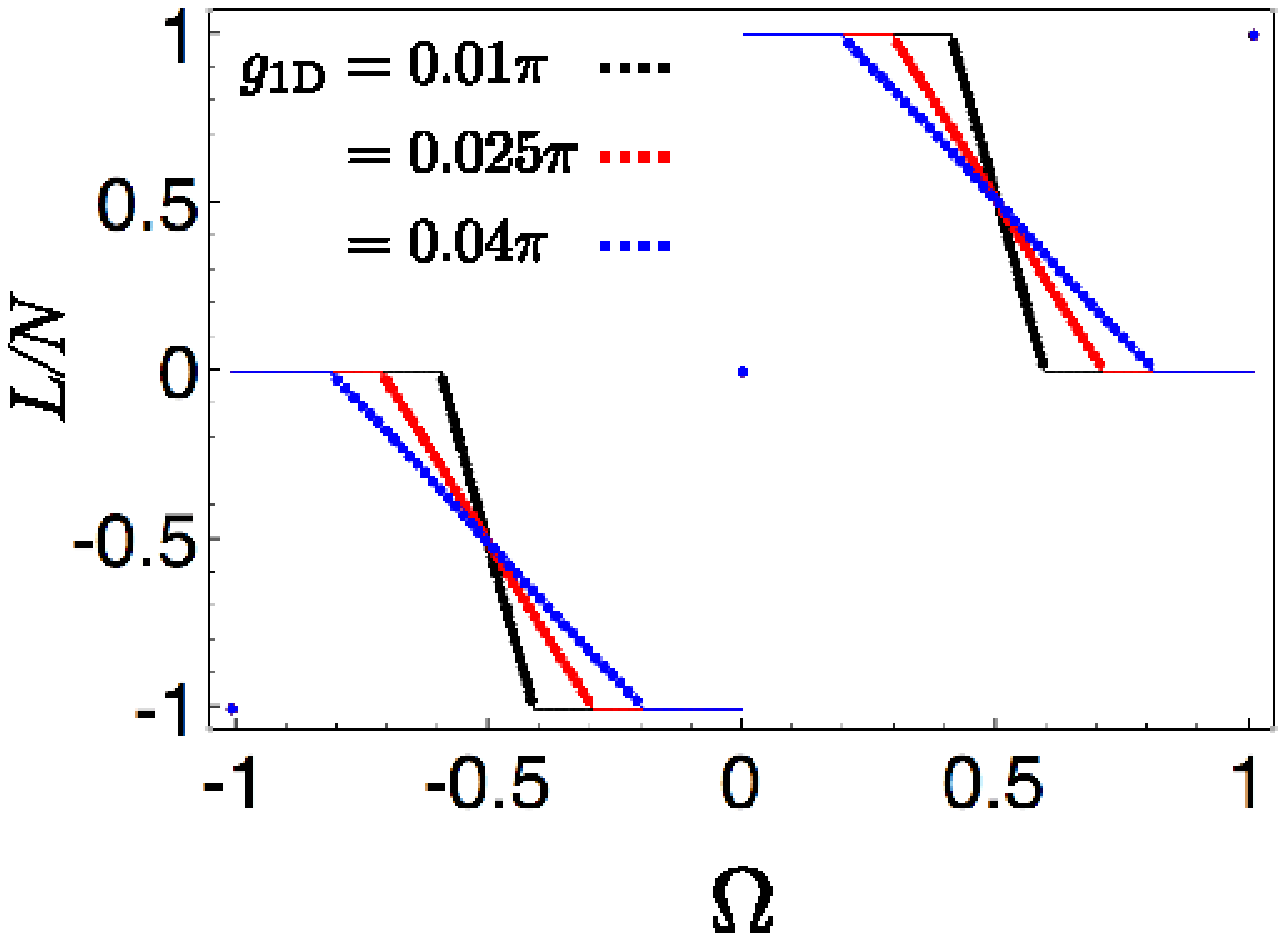}
\vspace*{-0.5cm}\end{center}
\caption{
The mean-field solutions can be extracted from our microscopic many-body analysis by using an extremization condition.  (a) Energy. The thin curves have integer average angular momenta and agree with the energy of the mean-field uniform-superflow states; the thick points signify noninteger angular momentum and have an energy close to that of the mean-field dark-soliton states.  (b) Angular momentum.  Note that we use the same parameter set as Figs.~\ref{fig1} and~\ref{fig2}.  Reproduced from~\cite{10:KCU}.
}
\label{fig3}
\end{figure}

The same argument applies to our case.
In the presence of any kind of noise, such as an infinitesimal distortion of
the trapping potential, quantum measurement of the matter wave, or whatever else
breaks the translation symmetry of the ring trap,
the realizable stationary state or metastable stationary
state is determined by extremization with respect to variations in angular momentum.
In order to find the metastable states we impose the condition
\begin{equation}\label{derE}
\partial_L E_{N,L}(\Omega)=0,
\end{equation}
with $\Omega$ and $g_{\rm 1D}$ held fixed.
Figure~\ref{fig3}(a) plots energy eigenvalues that satisfy this condition
as a function of $\Omega$; and Fig.~\ref{fig3}(b) shows the corresponding angular momentum.
These figures are quite similar to those given by mean-field theory, i.e.,
by imposing the stationary condition~(\ref{derE}) for the manifold of eigenvalues
we identify the mean-field stationary branches.

Referring to Fig.~\ref{fig3}(a), we cannot call the thick curve a soliton
branch in a rigorous sense, because each point is
an eigenvalue of the rLLH and thus the associated eigenstate still possesses translational symmetry,
unlike mean-field dark solitons.
Instead, we call all the angular-momentum states inside the swallowtail in
Fig.~\ref{fig2} the {\it soliton components}, because in the presence of
infinitesimal noise these states do form a broken-symmetry state, which we denote $|\chi\rangle$.
Soliton solutions of the NLSE can be interpreted in terms
of the eigensolutions of the rLLH as a state where the several
eigenvalues in the swallowtail region are collectively superimposed.

The energy associated with this superposition does not change significantly
because the energy required to make it is on the order of $1/N$. As a result,
the energy of the broken-symmetry soliton state $|\chi\rangle$ is also well
approximated by the thick curve in Fig.~\ref{fig3}(a).
In the presence of an infinitesimal symmetry-breaking potential, the angular momentum
is no longer a good quantum number. However, the expectation value of the angular momentum
$\langle \chi |\hat{L}|\chi \rangle$ agrees well with that of the dark solitons obtained by mean-field theory, and thus behaves like that shown in Fig.~\ref{fig3}(b)~\cite{08:KCU}.

With all these caveats in mind, we briefly state that the branch drawn by the thick curve
in Fig.~\ref{fig3}(a) is the {\it quantum soliton} branch in the weakly interacting regime.

We also calculated the second derivative $d^2 E_{N,L}(\Omega)/dL^2$ with respect to
$\Omega$ in order to check whether the metastable angular-momentum state
is a local maximum or minimum.
The superflow state with a CM quantum number
$J_0=\lfloor \Omega+1/2 \rfloor$ is indeed the ground state because
the second derivative is positive at that point, while the thick points
are local maxima with respect to $L$, since the second derivative is negative.

Let us summarize our first foray into the many-body theory.  In this section we obtained the yrast states $|N,L\rangle$ of the rLLH
by exact diagonalization.  We used an extremization condition to extract the mean-field results, among them the characteristic swallowtail shape in the spectrum.
We again found two kinds of metastable branches, uniform superflow and quantum soliton, consistent with mean-field theory.
The region of a high density of states, where the different angular-momentum states in the quantum theory densely
cross, indeed agrees with the soliton regime predicted by the mean-field and Bogoliubov theories.
The phrase ``quantum flesh sewn onto classical bones''
has been used elsewhere~\cite{davisED2004} as a visual metaphor, perhaps
inspired by x-ray images, to describe this accord.


\section{Strongly-Interacting Many-Body Theory: Tonks-Girardeau Limit}\label{BFmap}\index{Tonks-Girardeau gas}
\label{sec:TG}

Let us now move to the other extreme from the mean field.  Taking the interactions to be very large, the bosons become impenetrable and thus can be mapped to spinless fermions.  This is the TG regime.  We again solve the yrast eigenproblem and study the consequences.  In particular, we introduce the particle and hole excitations, which
are well defined in the fermionized gas, and show that these excitations are related to the mean-field stationary states in the opposite weakly interacting limit.

To be precise, the TG Bose-Fermi mapping theorem~\cite{60:Gir} is
\begin{equation}\label{BFM}
\Psi_B(\{\theta\})=\prod_{i > j}{\rm sgn}(\theta_i-\theta_j)\Psi_F(\{\theta\}),
\end{equation}
where $\Psi_B(\{\theta\})$ is the bosonic many-body wavefunction, expressed in first quantization, and $\Psi_F(\{\theta\})$ is the equivalent spinless fermionic many-body wavefunction.  This theorem holds for all the eigensolutions~\cite{02:GW}, and hence significantly simplifies
our eigenproblem.\footnote[4]{For simplicity of notation we will show the analytic expression only for an odd total
number of particles.
For an even number of particles the periodic boundary condition must be
taken as antisymmetric.}

\begin{figure}[!t]
\begin{center}
\includegraphics[width=1.0\textwidth]{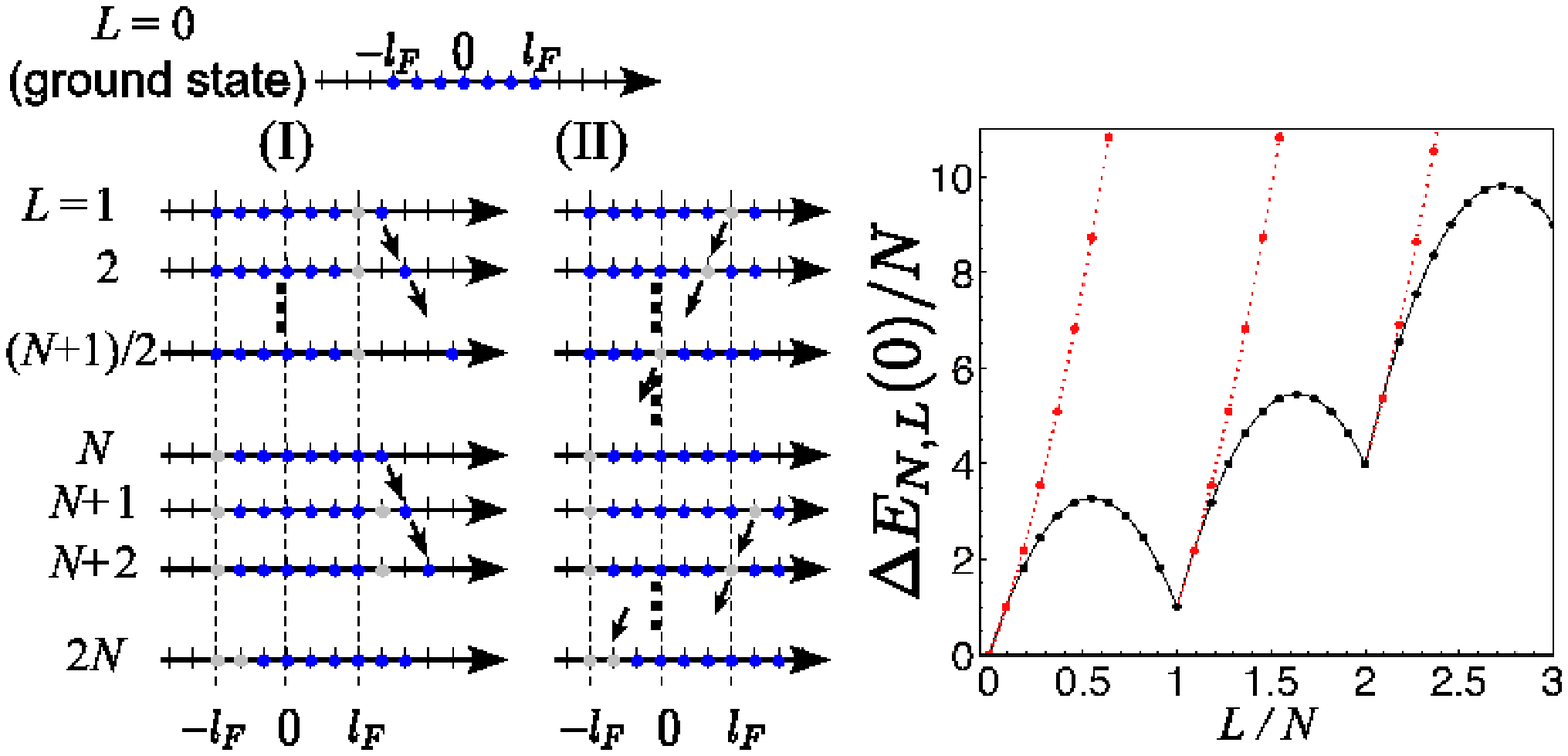}
\vspace*{-0.5cm}\end{center}
\caption{
(Left) A sketch of how to construct TG yrast states.
(I) Particle excitations where the angular momentum of a particle increases
while a hole is positioned at $l_F+J$.
(II) Hole excitations where a particle is placed at the lowest unoccupied
state and the angular momentum of a hole decreases.
When $L$ is an integer multiple of $N$, only the center-of-mass angular momentum
is shifted from the ground-state configuration.
(Right) Type I and type II excitation energies as a function
of $L/N$ for $N=11$ free fermions.  Reproduced from~\cite{10:KCU}.}
\label{fig5}
\end{figure}

The ground state of $N$ (odd) free fermions is obtained by the occupation of
the angular-momentum states from $l=-l_F$ to $l=l_F$ [see $L=0$ in Fig.~\ref{fig5}],
where $l_F\equiv (N-1)/2$ is the Fermi momentum. The ground-state energy is thus
\begin{equation}\label{TG_gs}
E_{N,L=0}(\Omega=0)=\textstyle\sum_{l=-l_F}^{l_F} l^2 =N(N^2-1)/12.
\end{equation}
Beyond the ground state, Lieb has shown~\cite{63:L} that excitation of the
repulsively interacting Bose gas in the thermodynamic limit has two branches.
The first branch is called \textit{type I}
and was shown to be in agreement with the Bogoliubov spectrum of plane waves in
the weakly-interacting regime.
The second branch is called \textit{type II}, and this was supposed
to be absent in the Bogoliubov spectrum.
For the fermionic formulation of the TG gas we observe that the type I and II branches correspond to particle
and hole excitations, respectively.

Let us reconsider these branches in the context of yrast states.
For the excited state $E_{N,L}(0)$ we can find the particle and hole excitations. The procedure is slightly technical, but we provide a sketch in Fig.~\ref{fig5}(a), and a brief description below.

\textit{Type I}: Remove a particle at the Fermi momentum $l_F$ and
place it at the momentum $l_F+L$.
For free fermions, there is no energy-level reconstruction in an
$(N\pm 1)$-particle system associated with removal or addition of a particle.
The energy of the type I excited state $E_{N,L}^{\rm (I)}(0)$ is
thus obtained as $E_{N,L}^{\rm (I)}(0)=E_{N,0}(0)-l_F^2+(l_F+L)^2$, or,
relative to the ground state,
\begin{equation}\label{exI}
\Delta E_{N,L}^{\rm (I)}(0) \equiv E_{N,L}^{\rm (I)}(0)-E_{N,0}(0)=L(N+L-1).
\end{equation}
There is no limitation on the single-particle angular momentum for Type I excitations.
Such excitations are doubly degenerate for $\Omega=0$, for $l \to -l$.
The excitations from the CMR states
are similarly obtained: to get the excitations from the $J=JN$ state, remove
a particle at $l_F+J$ and replace it at $l=l_F+J+L-JN$.  The resulting excitation energy is given by $\Delta E_{N,L}^{\rm (I)}(0) = J^2N+(l_F+J+L-JN)^2-(l_F+J)^2,\quad L \ge JN$.

\textit{Type II}: Starting from the ground state, remove a particle (create a hole)
at the momentum $l_F-L+1$ and place the particle at $l_F+1$, where
$0 \le L \le N$.  It is clear from Fig.~\ref{fig5} that the hole with this kind of
low-lying excitation energy be created only within the range $-l_F \le l \le l_F$.
The energy of this excited state is given by $E_{N,L}^{\rm (II)}(0)=E_{N,L}(0)-(l_F-L+1)^2+(l_F+1)^2$,
or, relative to the ground state,
$\Delta E_{N,L}^{\rm (II)}(0)=L(N-L+1)$, $0 < L \le N$.
Starting from the CMR state $L=JN$, we can extend this procedure to $L>N$, producing the series of humps shown in Fig.~\ref{fig5}.
From the fact that the excitation energy of the CMR state $L=JN$ is $\Delta E_{N,L=JN}(0)=J^2N$,
we can obtain the type II hole excitation energy for the general case as
\begin{equation}
\label{TG_exc}
\Delta E_{N,L}^{\rm (II)}(0) =J^2 N + (l_F+J+1)^2 -(l_F-L+JN+J+1)^2,
\end{equation}
where $JN < L \le (J+1)N$ and $\Delta E_{N,L}^{\rm (II)}(0) \equiv E_{N,L}^{\rm (II)}(0)-E_{N,0}(0)$ is again the energy relative to the ground state.
We note that Lieb's original study focused on the region $0 \le L/N \le 0.5$~\cite{63:L}.

\begin{figure}[t]
\begin{center}
\includegraphics[width=0.6\textwidth]{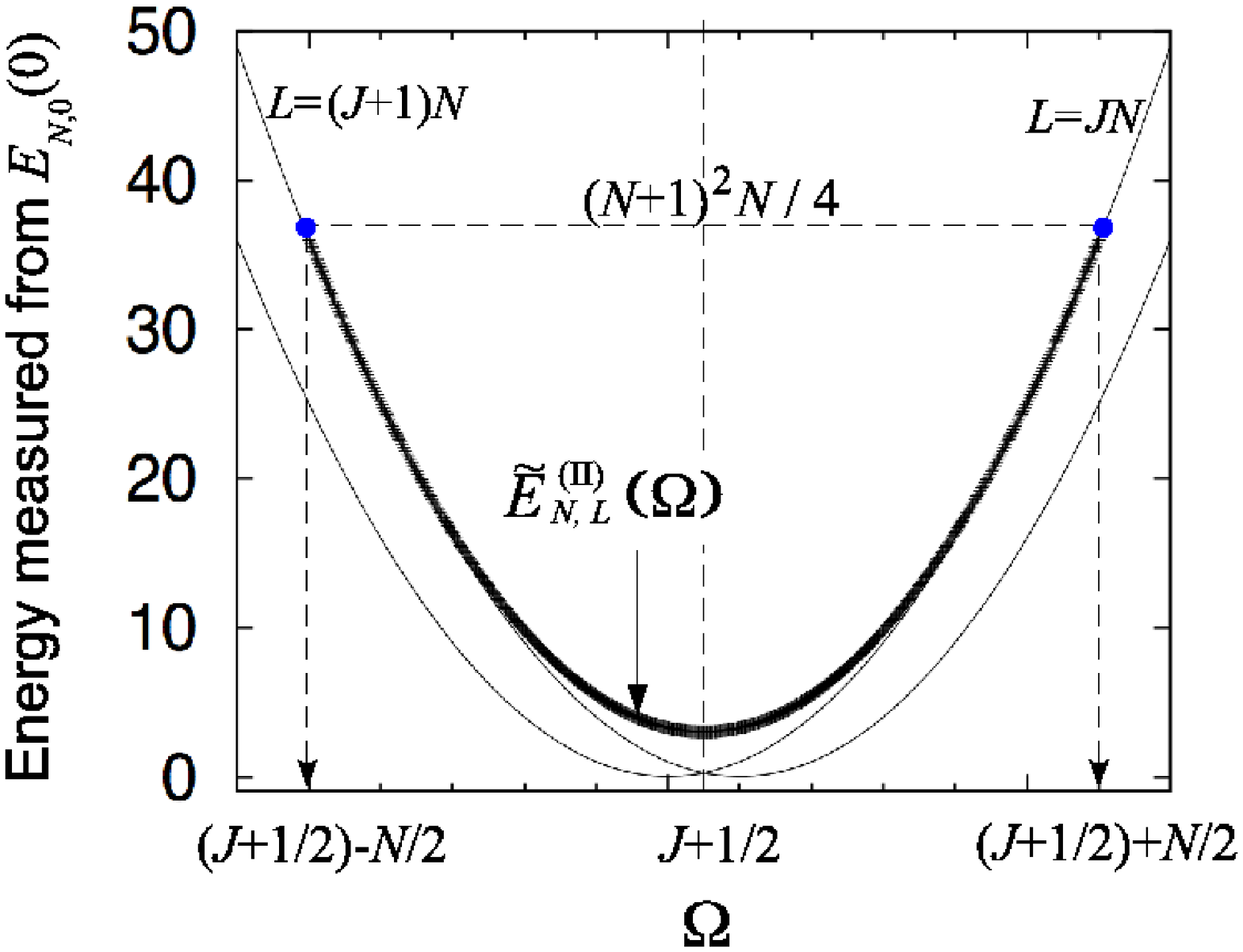}
\vspace*{-0.5cm}\end{center}
\caption{
We can again use an extremization condition to get the realizable yrast eigenstates in the strongly-interacting TG limit. Shown is the result obtained from the TG Bose-Fermi mapping for $N=11$ bosons.
All energies are plotted relative to the nonrotated ground-state energy.
The thin curves show the neighboring CMR states with angular momentum $L=JN$
and $L=(J+1)N$. The bold curve shows the intervening type II states
with angular momenta between these values.
The angular momentum decreases from $(J+1)N$ to $JN$ according to Eq.~(\ref{TG_stat_am})
along the bold curve.  The arrows show the critical frequencies for the QPT.  Reproduced from~\cite{10:KCU}.
}\label{fig6}
\end{figure}

Next we rotate these states to the TG yrast states for the rLLH, again using our Legendre transform.
The energy of the type II excited state for $JN < L \le (J+1)N$ measured relative to $E_{N,L=0}(0)$ is given by
\begin{equation}
\tilde{E}_{N,L}^{\rm (II)}(\Omega)=\Delta E_{N,L}^{\rm (II)}(0)-2\Omega L + \Omega^2 N\,.
\end{equation}
We can again get the key solutions from the extremization condition~(\ref{derE}), where now we take the energy to be that of type II excitations.
By inspection, CMR states $L=JN$ are metastable states, either ground or excited.
Condition~(\ref{derE}) gives another metastable angular momentum,
\begin{equation}\label{TG_stat_am}
\bar{L}=(N+1)(J+1/2)-\Omega,
\end{equation}
and the corresponding energy,
\begin{equation}
\label{TG_stat_ene}
\tilde{E}_{N,\bar{L}}^{\rm(II)}(\Omega)
=(N+1)[\Omega-(J+1/2)]^2+N(N+1)/4.
\end{equation}
As a function of $\Omega$ this is a parabolic curve, shown in Fig.~\ref{fig6}.

We again find certain critical angular frequencies where the metastable type II branch disappears and merges into the CMR branch:
\begin{equation}
\Omega_{\mathrm{cr}}^{\mp}=(J+1/2)\mp N/2.
\label{eqn:TGcritical}
\end{equation}
The stable angular momentum approaches $\bar{L}=(J+1)N$ at $\Omega_{\mathrm{ cr}}^{-}$
and $\bar{L}=JN$ at $\Omega_{\mathrm{cr}}^{+}$, respectively, and the corresponding energy
coincides with the energy of CMR states.  Equation~(\ref{eqn:TGcritical}) determines the critical boundary for our QPT in the strongly interacting limit.

This whole picture closely matches that of the uniform-superflow to dark-soliton transition in the weakly interacting limit,
where there exists a critical angular frequency at which the soliton branch bifurcates
from the superflow branch: here type II yrast states branch off of the type I CMR states. We therefore interpret the hole excitations in the TG limit with the soliton branch in the weakly interacting limit.  It remains to fill in the picture between, as we proceed to do in the following section via the Bethe equations.

\section{Bridging All Regimes: Finite-Size Bethe Ansatz}\label{sec:BA}\index{Bethe ansatz}

\begin{figure}
\begin{center}
\includegraphics[width=0.6\textwidth]{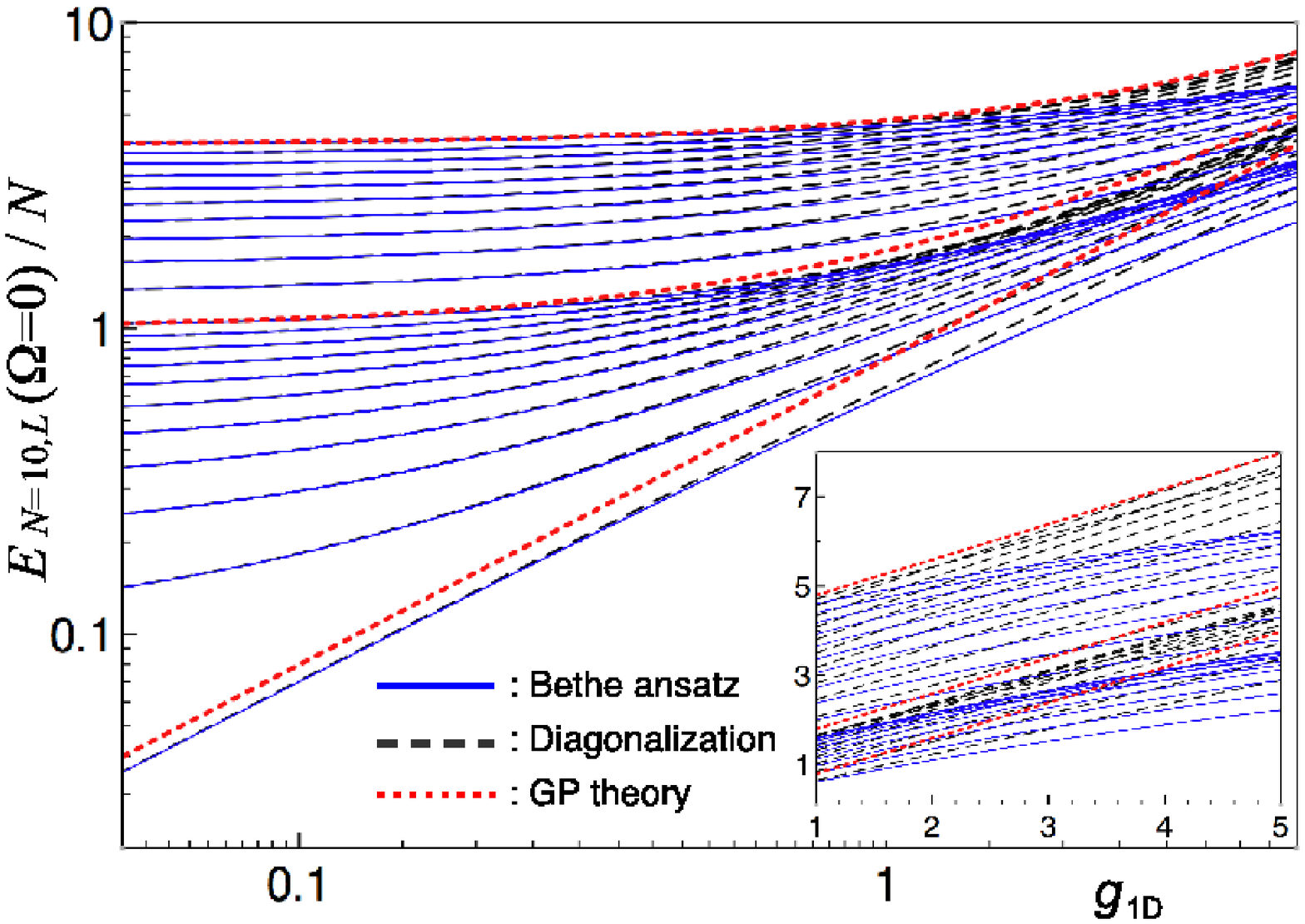}
\vspace*{-0.5cm}\end{center}
\caption{
Evaluation of three many-body solution methods, Bethe equations, exact diagonalization, and mean-field NLSE (labeled GP Theory), in the weakly- to medium-interacting regime in terms of yrast eigenenergies (note log scale).  We take $N=10$ particles and CMR quantum number $J\in\{0,1,2\}$ only for the purposes of illustration.  The inset enlarges the medium-interacting regime on a linear scale. Reproduced from~\cite{10:KCU}. }\label{fig7}
\end{figure}

In this section we will show that our QPT stretches over the whole interaction range, and also validate our physical interpretation of Lieb Type II excitations.
We note that, complementary to our work, the spectrum of the LLH has been obtained~\cite{06:CB} by treating the
inverse of the TG parameter, which is infinite at the TG regime, as the expansion
parameter, and its analytical interpolation was given recently~\cite{09:CB}.

Using the Bethe ansatz, we obtain the $N$ simultaneous \textit{Bethe equations} for our system,
\begin{equation}\label{bethe_eq}
(-1)^{N-1}e^{-2\pi i \ell_n}=
\prod_{m=1}^N \frac{\ell_n-\ell_m
+i g_{\rm 1D}/2}{\ell_n-\ell_m- i g_{\rm 1D}/2},
\end{equation}
which determine the set of values $\{ \ell_n \}$, called \textit{quasi-angular momenta} for each atom $n\in\{1,\ldots,N\}$.\footnote[5]{It is sufficient to solve for only the real part of these equations~\cite{10:KCU}.}  The quasi-angular momenta fully characterize the angular momentum and energy, given by $L=\sum_{n=1}^N \ell_n$ and $E_{N,L}(\Omega=0)=\sum_{n=1}^N \ell_n^{2}$, respectively.

We numerically solve the real part of the Bethe equations~(\ref{bethe_eq}) for
each set of energy levels
characterized by the different total angular momenta.
The numerical solution of Eqs.~(\ref{bethe_eq}) is highly sensitive to the initial
set of trial values of $\{\ell_n\}$.
If this initial set is sufficiently close to a solution for a target angular-momentum state,
the set of solutions $\{\ell_n\}$ can be correctly obtained.  In contrast, if the initial set
is closer to another angular-momentum state, the total angular momentum
reveals undesired jumps, deviating from the target angular momentum. In such a case we again start from
another initial set of trial values of quasi-momenta.  Our algorithm is outlined in~\cite{10:KCU}.

In Fig~\ref{fig7}, we first perform a comparative study for mean-field, TG, and Bethe solution methods in the weakly- to medium-interacting regime.\footnote[6]{We note that the concept of yrast state for the angular momenta $JN < L < (J+1)N$
does not exist in the mean-field theory: this theory is concerned only with
the single-particle angular momentum, which coincides with the average angular momentum
in this theory. We thus plot the mean-field energy for the integral single-particle angular momenta.}
As can be seen, for $g_{\rm 1D} \lesssim \mathcal{O}(1)$, the three methods agree to a few percent.

\begin{figure}[!t]
\begin{center}
\includegraphics[width=1.0\textwidth]{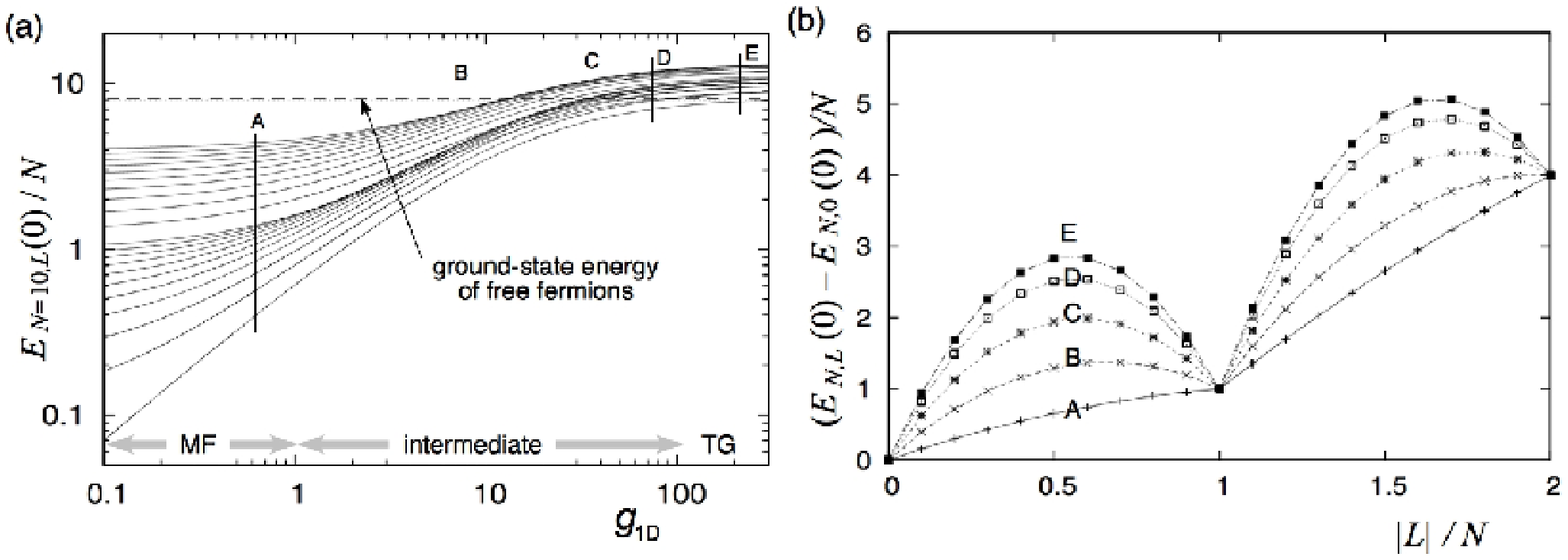}
\vspace*{-0.5cm}\end{center}
\caption{
(a) Solutions of the Bethe equations in the nonrotating frame for a range of angular momenta $|L| \le 2N$ and $N=10$ particles.  The horizontal dashed line corresponds to the energy of
free fermions (\ref{TG_gs}) with the same number of atoms.
(b) Excitation energies of yrast states as a function
of average angular momentum $|L|/N$ for fixed strengths of interaction.
A, B, C, D, and E are marked in (a).  The kinks are again the CMR states.  Reproduced from~\cite{10:KCU}.
}\label{fig8}
\end{figure}

In Fig.~\ref{fig8}(a) we show the spectra over the entire interaction range, for the system in the nonrotating frame: specifically, $E_{N,L}(\Omega=0)$ for $|L|\in\{0,\pm 1,\ldots, \pm 2N\}$.  All ground- and excited-state energies monotonically increase with respect
to $g_{\rm 1D}>0$.  However, the energy does not monotonically increase with respect to
the total angular momentum for a fixed strength of interaction. This is illustrated in Fig.~\ref{fig8}(b).  The labels A, B, C, D, and E correspond
to the cross-cuts labeled in panel (a).
While in the very weakly-interacting limit (curve A) the spectrum still looks almost linear,
as the interaction increases, the kinks in the yrast spectra at the location of CMR state,
$L=JN$, become more pronounced due to the large increase in the energy
of the yrast states with $L$ in between neighboring CMR states at $JN$ and $(J+1)N$. For strong interactions (curve E),
the system is in the TG regime; compare to the TG calculation for Type II excitations from Fig.~\ref{fig5}(b).
We observe numerically that
the excitation energy of the CMR state $L=JN$ is independent of $g_{\rm 1D}$, and is given
by Eq.~(\ref{sf_ex_ene}), namely $(E_{N,L}-E_{N,0})/N=(L/N)^2$.
This follows from the nature of the CMR state $L=JN$,
which is just a Galilean boost of the nonrotating state;
under this transformation interactions are unchanged.

\begin{figure}[!t]
\begin{center}
\includegraphics[width=0.6\textwidth]{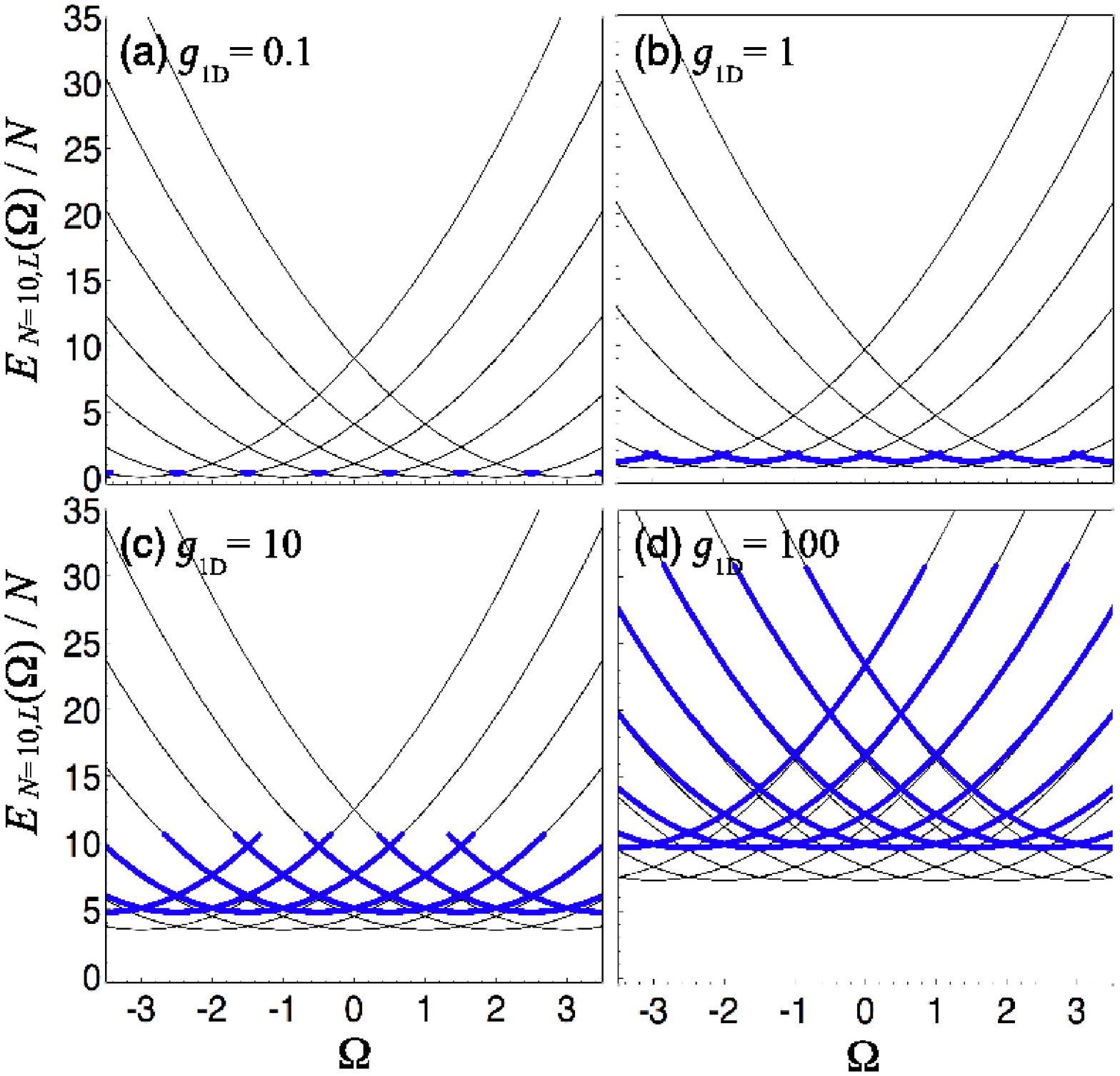}
\vspace*{-0.5cm}\end{center}
\caption{
Spectra of the rLLH obtained via the Bethe equations and satisfying the extremization condition for (a) weakly-interacting regime,
(b) and (c) medium-interacting regime, and (d) strong-interacting regime.
The thin curves show CMR states, while the thick curves show yrast states with $L/N$ noninteger, and come
from the type II excitation branch.  Reproduced from~\cite{10:KCU}.
}\label{fig9}
\end{figure}

We proceed to the rotating frame, as before, via our Legendre transform, which can be framed in terms of quasi-angular momenta as
\begin{equation}\label{ene_rot}
E_{N,L}(\Omega)=\textstyle\sum_{j=1}^N (\ell_j-\Omega)^2
= E_{N,L}(0) -2\Omega L +\Omega^2N.
\end{equation}
The results are shown in Fig.~\ref{fig9} for various strengths of interaction.
The thin curves are parabolas $(\Omega-J)^2+V_{\rm int}$ for various values of
center-of-mass quantum numbers $J$.
The lowest possible energy of the CMR state is thus given by $V_{\rm int}$ at $\Omega=J \in \{\mathbb{Z}\}$.
The thick curves plot other stable yrast states of noninteger $L/N$ from the condition~(\ref{derE}).
The weakly-interacting mean-field regime is shown in Fig.~\ref{fig9}(a), where
the type II branch that satisfies the metastable condition just starts to appear.
Thus these are the energies of the quantum solitons [see also Fig.~\ref{fig3}(a)].
As the interaction increases [Figs.~\ref{fig9}(b) and \ref{fig9}(c)]
the domain with the swallowtail shape
enclosed by the two CMR branches, as well as the size of the type II branch, increases.
In the TG limit [\ref{fig9}(d)], the area of the swallowtail region saturates the spectra.

\begin{figure}[!t]
\begin{center}
\includegraphics[width=0.7\textwidth]{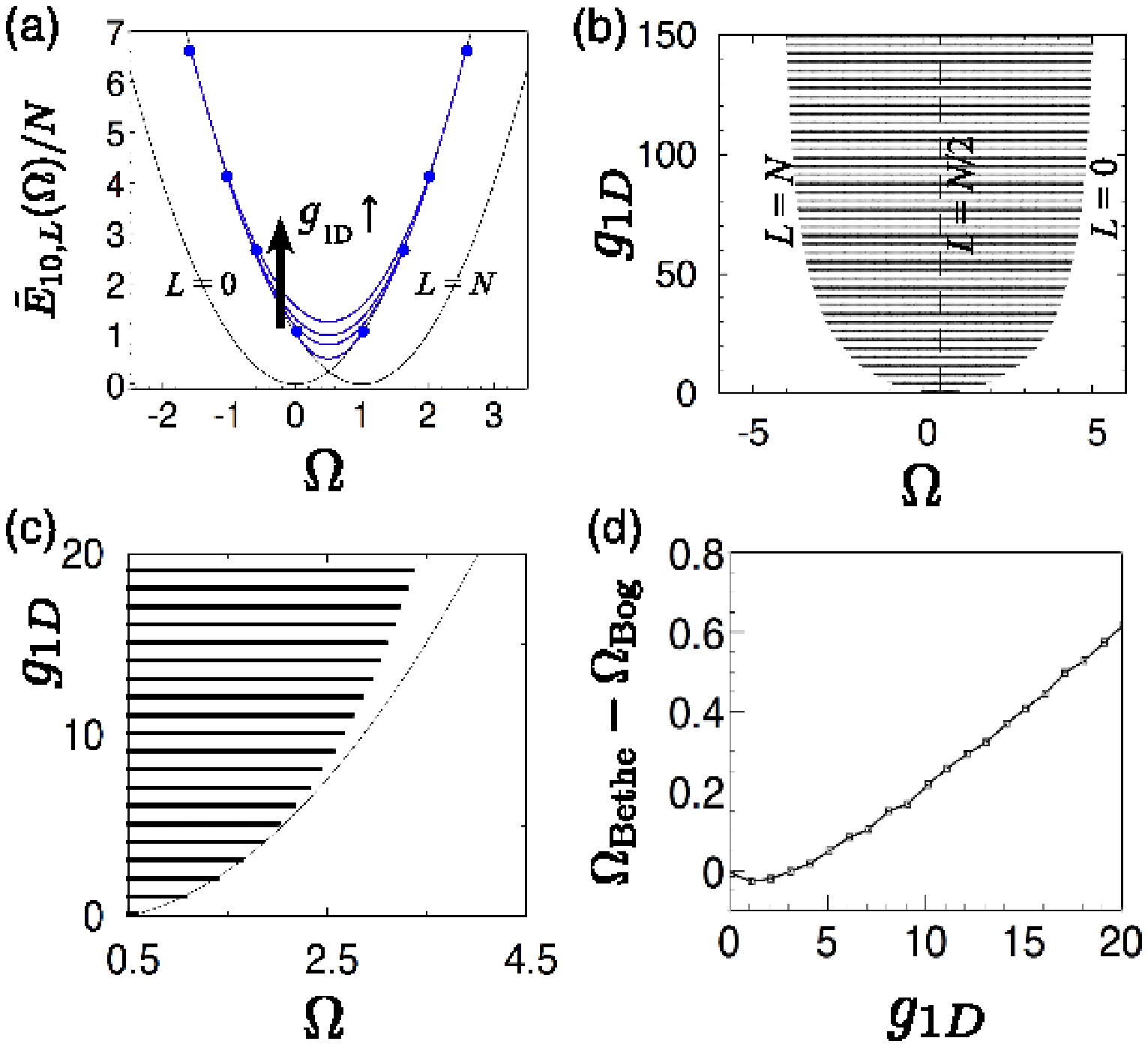}
\vspace*{-0.5cm}\end{center}
\caption{
(a) Two CMR states with $L=0,N$ (thin curves) and
the type II branch (thick curve) that connects them.
The energy is defined relative to the interaction energy $V_{\rm int}$.
(b) The quantum phase diagram over all interaction strengths; in the gray region yrast states corresponding to Lieb's type II excitations allow for a continuous change in angular momentum, while in the white region only CMR states with quantized $L/N$ are allowed.
(c) Enlargement of (b) in the weakly-interacting regime. The solid curve is
the critical boundary given by the Bogoliubov theory.
(d) Difference between the phase boundaries given by the Bethe ansatz and
the Bogoliubov theory. The Bogoliubov critical boundary overestimates the correct Bethe result.  Reproduced from~\cite{10:KCU}.
}\label{fig10}
\end{figure}

These behaviors are quantitatively summarized in Fig.~\ref{fig10}(a),
which shows the energy $E_{N,L}(\Omega)/N$ of metastable states relative to
the interaction energy $V_{\rm int}$ at each strength of interaction.
The CMR branches drawn by the thin curves no longer have a $g_{\rm 1D}$-dependence
because of the subtraction of $V_{\rm int}$,
while the thick curve gradually increases the domain over which it extends as the interaction increases.
For simplicity we plot only two CMR branches with angular momenta $L=0,N$,
and a metastable state associated with the type II branch that smoothly connects these two CMR states.

As $\Omega$ increases, the thick curve bifurcates from the CMR branch
with angular momentum $L=N$ at a certain critical $\Omega < 0.5$, and
at $\Omega=0.5$ the energy is minimal.
As $\Omega$ increases further, this branch
smoothly merges into the CMR branch with angular momentum $L=0$
and eventually disappears at a certain critical $\Omega > 0.5$.
We therefore find that the same kind of energy bifurcation which
was found in the mean-field theory
persists over the full range of repulsive interactions.
Figure~\ref{fig10}(b) illustrates this idea by plotting the existence range of the metastable-state
type II excitation branch.
The shaded area indicates the existence of such a branch.
The angular momentum on the lower critical boundary is given by $L=N$, and
the angular momentum linearly decreases as $\Omega$ increases, just like in Fig.~\ref{fig3}(b).
At $\Omega=0.5$ (vertical dashed line), the value of the angular momentum is
given by $L=N/2$ irrespective of the strength of interaction.
At a certain value of $\Omega (> 0.5)$ the angular momentum eventually
goes to zero, causing the metastable hole excitation branch to disappear.
This behavior corresponds to the fact that in the mean-field theory
the type II branch bifurcates from the uniform-superflow regime,
developing nodes, and it again merges into the uniform-superflow regime
with the increase of $\Omega$~\cite{08:KCU, 09:KCU}.
The critical boundary approaches $\Omega=(J+1/2)\pm N/2$ in the
strongly-interacting regime.

Finally, we explicitly evaluate the BdGE analytical predictions for the critical boundary.  In Fig.~\ref{fig10}(c)-(d) the Bethe-equation prediction for the critical boundary is compared with
that obtained from the BdGE in the weakly-interacting regime.
We observe that Bogoliubov theory predicts the quantitatively correct
critical boundary to the 5$\%$ level up to $g_{\rm 1D} \lesssim 5$ (for $N=10$),
but it significantly overestimates the boundary as the interaction increases.

Thus we find a critical boundary for all values of interaction strength.


\section{Conclusions and Outlook}\label{sec:conclusion}

We have tied together a number of outstanding issues in ultracold quantum gases.  The first is the understanding of quantum effects on characteristic mean-field solutions, namely dark solitons.  The second is the concept of a phase transition in a mesoscopic system which is strongly isolated from its environment.  Other issues we addressed included the physical interpretation of Lieb's Type-II excitations and the correspondence of various 1D techniques, ranging from the NLSE to the Bethe equations to the TG mapping.

The first issue has occupied many people over the last decade.  Work in this area includes quantum delocalization of dark solitons in the weakly interacting regime~\cite{03:DKS,martinAD2009}, decay of dark solitons in the more strongly interacting regime~\cite{carr2009}, the search for dark solitons in the extreme limit of the TG gas via the TG Bose-Fermi mapping~\cite{girardeau2000b}, and most recently the role of thermal fluctuations in defect formation by the Kibble-Zurek mechanism~\cite{damski2009}.  Our results indicate that while the mean-field dark soliton bears some meaning in the weakly-interacting regime, where one can make a close correspondence between the underlying many-body quantum soliton branch and the symmetry-broken mean-field soliton, in the medium- to strongly-interacting regime no such correspondence is likely.  Thus for stronger interactions we do not expect to observe a density notch with a characteristic phase structure.  Given recent experiments tuning the interactions over seven orders of magnitude and the increasing commonness of ring traps, we hope to see tests of these ideas in the near future.  A new dynamical study by Brand and Kolovsky~\cite{10:BK} indicates that in various regimes stretching from the NLSE to the TG gas our ideas have some merit; after all, if our predicted many-body generalization of dark solitons took the lifetime of the universe to nucleate, our ideas would remain restricted to the realm of theory; this appears not to be the case.  A complementary study to Brand and Kolovsky's might use MPS methods to address this question.

The second issue is trickier.  We all have a thermodynamic view of phase transitions and states of matter drummed into us, and new contexts such as graphene are showing that some of these ideas have to be generalized.  In our case we find a QPT, technically a crossover, that nevertheless displays characteristics of phase transitions: topological properties of the system change and the energy reveals a cusp at some level of derivative.  Moreover, our QPT is inherently finite size and has no meaning in the thermodynamic limit.
For isolated systems such as ultracold quantum gases or nuclei (see Chap.~27) a QPT in excited states is a useful concept, in contrast to QPTs in solid-state materials.  The mean-field theory is expressed in a nonlinear partial differential equation, the NLSE; it is useful to note that bifurcations appearing in the solution space of such effective nonlinear theories often point to a QPT in the underlying linear many-body theory.

As a very brief summary of other issues addressed in this chapter, we showed that the soliton-train uniform-superflow picture falls naturally out of consideration of center-of-mass-rotation states and quantum soliton states, the latter being yrast states contained within a swallowtail.  These considerations provided a complete description of the finite-system metastable QPT from the weakly- to strongly-interacting regimes.  In the process we cleared up the long-standing problem of the physical interpretation of Lieb's type-II or hole excitation branch: type-II excitations are exactly the yrast states, which in the weakly-interacting limit can be expressed as dark-soliton trains, appearing in broken-symmetry form in mean-field theory.

Our static study can be extended to many different contexts in ultracold quantum gases, for example, spinor Bose systems, where the hyperfine structure of the constituent atoms in the BEC plays the role of a spin, and one finds spin-one or even spin-two and higher models, as described in Chap.~10.  A number of metastable and/or finite-size QPTs are possible in these systems.  Ultracold fermions also provide a fine candidate, and we expect a metastable QPT occurs in the Bardeen-Cooper-Schrieffer (BCS) superconducting phase.  One could trace this QPT through the  BCS-to-BEC transition.  This problem is particularly interesting halfway through the crossover in the unitary regime,\footnote[7]{The unitary regime is where the scattering length for binary contact interactions, $a_s$, diverges: $k_F a_s \to \infty$, with $k_F$ the Fermi momentum.} a many-body system under intensive investigation in ultracold quantum gases, in the quark gluon plasma (see Chap.~25), and in the AdS/CFT (anti-de Sitter / conformal field theory) mapping (see Chap.~28).

More generally, we can ask if the concept of QPTs in finite systems is useful beyond ultracold quantum gases.  What happens to critical exponents?  Can they be defined without the use of finite-size scaling?  Is the concept of a universality class useful?  Can backing off from the thermodynamic limit but retaining the QPT concept in some fashion help us better understand the fundamental physics of mesoscopic systems and help us invent better nano-devices?

\textit{Acknowledgments} -- This work was supported by the National Science Foundation under Grant PHY-0547845 as part of the NSF CAREER program (L.D.C.), a Grant-in-Aid for Scientific Research under Grant Numbers 17071005 (M.U.) and 21710098 (R.K.), and by the Aspen Center for Physics.


\begin{thebibliography}{50}

\bibitem{leggett2001}
A. J. Leggett, Rev. Mod. Phys. {\bf 73}, 307 (2001).

\bibitem{09:Hulet}
S. E. Pollack \textit{et al.}, Phys. Rev. Lett. {\bf 102}, 090402 (2009).

\bibitem{63:L}
E. H.~Lieb, Phys. Rev. {\bf 130}, 1616 (1963).

\bibitem{63:LL}
E. H.~Lieb and W.~Liniger, Phys. Rev. {\bf 130}, 1605 (1963).

\bibitem{03:Dem}
S. O.~Demokritov \textit{et al.}, Nature {\bf 426}, 159 (2003); S.~Gupta \textit{et al.},
Phys. Rev. Lett. {\bf 95}, 143201 (2005); A. S.~Arnold, C. S. Garvie, and E. Riis,
Phys. Rev. A {\bf 73}, 041606(R) (2006); S. R.~Muniz \textit{et al.},
Opt. Express {\bf 14}, 8947 (2006); C.~Ryu \textit{et al.}, Phys. Rev. Lett. {\bf 99}, 260401 (2007); B. P.~Anderson, K.~Dholakia, and E. M.~Wright,
Phys. Rev. A {\bf 67}, 033601 (2003).

\bibitem{leggett1999}
A. J. Leggett, Rev. Mod. Phys. {\bf 71}, S318 (1999).

\bibitem{penrose1956}
O. Penrose and L. Onsager, Phys. Rev. {\bf 104}, 576 (1956).

\bibitem{burger1999}
S. Burger \textit{et al.}, Phys. Rev. Lett. {\bf 83}, 5198 (1999).

\bibitem{denschlag2000}
J. Denschlag \textit{et al.}, Science {\bf 287}, 97 (2000).

\bibitem{weller2008}
A. Weller \textit{et al.}, Phys. Rev. Lett. {\bf 101}, 130401 (2008).

\bibitem{stellmerS2008}
S. Stellmer \textit{et al.}, Phys. Rev. Lett. {\bf 101}, 120406 (2008).

\bibitem{dunjko2001}
V. Dunjko, V. Lorent, and M. Olshanii, Phys. Rev. Lett. {\bf 86}, 5413 (2001).

\bibitem{jackson2006}
B. Jackson, C.~F. Barenghi and N.~P. Proukakis, J. Low Temp. Phys. {\bf 148}, 387 (2006).

\bibitem{03:DKS}
J. Dziarmaga, Z.~P. Karkuszewski, and K. Sacha, J. Phys. B: At. Mol. Opt. Phys. {\bf 36}, 1217 (2003).

\bibitem{carr2009}
R. V. Mishmash and L. D. Carr, Phys. Rev. Lett. {\bf 103}, 140403 (2009);
R. V. Mishmash, I. Danshita, C. W. Clark, and L. D. Carr, Phys. Rev. A {\bf 80}, 053612 (2009).

\bibitem{kivshar1998}
Y. S. Kivshar, Phys. Rep. {\bf 298}, 81 (1998).

\bibitem{gordon2}
J. P. Gordon and H. A. Haus, Opt. Lett. {\bf 11}, 665 (1986).

\bibitem{fedichev1999}
P. O. Fedichev, A. E. Muryshev, and G. V. Shlyapnikov, Phys. Rev. A {\bf 60}, 3220 (1999).

\bibitem{martinAD2009}
A. D. Martin and J. Ruostekoski, Phys. Rev. Lett. {\bf 104}, 194102 (2010).

\bibitem{deuar2010}
P. Deuar, J. Dziamarga, and K. Sacha, 	Phys. Rev. Lett. {\bf 105}, 018903  (2010); R. V. Mishmash and L. D. Carr, Phys. Rev. Lett. {\bf 105}, 018904 (2010).

\bibitem{damski2009}
B. Damski and W. H. Zurek, 	Phys. Rev. Lett. {\bf 104}, 160404 (2010).

\bibitem{99:BR}
D. A.~Butts and D. S.~Rokhsar,
Nature {\bf 397}, 327 (1999).

\bibitem{99:BP}
G. F.~Bertsch, T.~Papenbrock,
Phys. Rev. Lett. {\bf 83}, 5412 (1999).

\bibitem{80:IT}
M.~Ishikawa and H.~Takayama,
J. Phys. Soc. Jpn. {\bf 49}, 1242 (1980).

\bibitem{yukalov2005}
V. I. Yukalov and M. Girardeau, Laser Phys. Lett. {\bf 2}, 375 (2005).

\bibitem{geim2007}
A. K. Geim and K. S. Novoselov, Nature Materials {\bf 6}, 183 (2007).

\bibitem{04:KWW}
T.~Kinoshita, T.~Wenger, and D.~Weiss, Science {\bf 305}, 1125 (2004).

\bibitem{08:KCU}
R.~Kanamoto, L. D.~Carr, and M.~Ueda, Phys. Rev. Lett. {\bf 100}, 060401 (2008).

\bibitem{09:KCU}
R.~Kanamoto, L. D.~Carr, and M.~Ueda, Phys. Rev. A {\bf 79}, 063616 (2009).

\bibitem{10:KCU}
R.~Kanamoto, L. D.~Carr, and M.~Ueda, Phys. Rev. A {\bf 81}, 023625 (2010).

\bibitem{05:KSU}
R.~Kanamoto, H.~Saito, and M~.Ueda,
Phys. Rev. Lett. {\bf 94}, 090404 (2005).

\bibitem{00:CCR}
L. D.~Carr,  C. W.~Clark, and W. P.~Reinhardt,
Phys. Rev. A {\bf 62}, 063610 (2000).

\bibitem{carr2000e}
L. D.~Carr,  M. A. Leung, and W. P.~Reinhardt,
J. Phys. B: At. Mol. Opt. Phys. {\bf 33}, 3983 (2000).

\bibitem{carr2001e}
L. D. Carr, J. Brand, S. Burger, and A. Sanpera,
Phys. Rev. A {\bf 63}, 051601(R) (2001).

\bibitem{muth2009}
D. Muth, B. Schmidt, and M. Fleischhauer, e-print arXiv:0910.1749 (2009).

\bibitem{98:Ols}
M.~Olshanii, Phys. Rev. Lett. {\bf 81}, 938 (1998).

\bibitem{01:FS}
A. L.~Fetter and A. A.~Svidzinsky,
J. Phys.: Condens. Matter {\bf 13}, R135 (2001); L. J.~Garay {\it et al.},
Phys. Rev. Lett. {\bf 85}, 4643 (2000).

\bibitem{60:Gir}
M.~Girardeau, J. Math. Phys. (N.Y.) {\bf 1}, 516 (1960).

\bibitem{73:Legg}
A. J.~Leggett, Phys. Fenn. {\bf 8}, 125 (1973).

\bibitem{13:Sag}
G.~Sagnac, C.R. Acad. Sci. {\bf 157}, 708 (1913); {\bf 157}, 1410 (1913);
J. Phys. {\bf 4}, 177 (1914).

\bibitem{davisED2004}
E. D.~Davis, Phys. Rev. A {\bf 70} 032101 (2004); Alisa Bokulich,
\emph{Reexamining the Quantum-Classical Relation: Beyond Reductionism and Pluralism}
(Cambridge University Press, New York, 2008).

\bibitem{02:GW}
M. D.~Girardeau and E. M.~Wright, Laser Phys. {\bf 12}, 8 (2002).

\bibitem{06:CB}
A. Y.~Cherny and J.~Brand,
Phys. Rev. A {\bf 73}, 023612 (2006)

\bibitem{09:CB}
A. Y.~Cherny and J.~Brand,
Phys. Rev. A {\bf 79}, 043607 (2009).

\bibitem{10:BK}
J. Brand and A. Kolovsky,
to be submitted (2010).

\bibitem{girardeau2000b}
M. D. Girardeau and E. M. Wright,
Phys. Rev. Lett. {\bf 84}, 5691 (2000).


\end{thebibliography}
\end{document}